\documentclass[a4,12pt,romanian,oneside]{report}

\usepackage[a4paper,left=35mm,right=25mm,bottom=35mm]{geometry}

\usepackage{fancyhdr}

\normalsize

\usepackage[dvips]{graphicx}
\usepackage{mathtools}
\usepackage{wrapfig}
\usepackage{xfrac}
\usepackage{color}
\usepackage{graphics}
\usepackage{float}
\usepackage{titlesec}
\usepackage{setspace}
\usepackage{sectsty}
\usepackage{amsmath,amssymb,lmodern}
\usepackage{subcaption}
\usepackage{fancyhdr}
\usepackage{url}

\chapterfont{\LARGE\centering}
\sectionfont{\large\centering}
\subsectionfont{\normalsize\centering}

\def\bq{\begin{eqnarray}}
\def\eq{\end{eqnarray}}
\def\nn{\nonumber}
\date{}

\large

\begin{document}
\onehalfspace
\def\contentsname{Contents}

\begin{titlepage}
 \ \vspace{2in} \\
 \huge{The response of a Unruh-deWitt particle \vspace{0.1in} \\  detector in a thin-shell wormhole spacetime} 
 \large{}
 
 \vspace{1in} \ \\
\begin{centering}
 {\bf \large Robert Blaga} \\
 \emph{\large Department of Physics, West University of Timi\c soara,} \\
 \emph{\large Bd. Vasile P\^ arvan 4, Timi\c{s}oara 300223, Romania} \\
\end{centering}

\end{titlepage}

\begin{abstract}
We investigate the transition probability of a Unruh-deWitt particle detector evolving in flat space and in a wormhole spacetime, in various scenarios. In Minkowski space, we look at the response of the detector on trajectories 
having discontinuities and rapid variations, as well as the effect of finite-time coupling. It is found that these features induce spurious oscillations in the probability and rate of transition. At large 
times the oscillations are damped and the probability tends to a constant value. Next, we look at the response of an inertial detector on a radial trajectory that passes through a thin-shell wormhole.
After finding the appropriate modes, we look at the renormalized detector response, defined by subtracting the flat space analogues from the partial probabilities. The resulting curve has a peak around
the wormhole throat followed by a period of damped oscillations, before stabilizing to a constant value. This is very similar to the flat space results, which is surprising given that in this case the trajectory is 
continuous. The features of the transition probability are due entirely to the nontrivial topology induced by the wormhole. 
\\ \\ \\ \\ \\ 
\large{The manuscript represents the translated and expanded on version of the author's MSc dissertation. The thesis in original form can be found at: \url{http://quasar.physics.uvt.ro/~cota/CCFT/pdfuri/RB_MSc_dissertation.pdf}}
\end{abstract}

\tableofcontents
\chapter*{Introduction}
\addcontentsline{toc}{chapter}{\ \ \ \ Introduction}
With the publication in 1974 of the landmark paper by S.~Hawking \cite{a5,a4}, the ideea that there is a link between phenomena at the interface of quantum field theory and general relativity, and thermodynamics
had started seeping into the mainstream. Hawking had shown that that an observer situated at inifinity would see a Black Hole (BH), formed through the gravitational collapse of a star, as emitting radiation with a thermal 
spectrum at temperature:
\begin{equation}
 T = \frac{\hbar \kappa}{2\pi ck},
\end{equation}
where $\kappa$ is the surface gravity of the BH. \par
In an effort to better understand Hawkings result, W.Unruh introduced in a paper published in 1976, the concept of particle detectors \cite{a6}. Such a detector is a thoeretical tool with the help of which the
state of a quantum field could be investigated. The benefit of such an approach is that shunts the need to define what a ``particle'' means, a task which in curved spacetimes can be problematic. Using particle detectors, the notion of particle
can be defined operationally as that ``something'' which the detector detects. The detector usually is a simple quantum mechanical system which can absorb or emit quanta of the quantum field, and hence suffer transitions between
its internal energy levels. Unruh showed that the interesting link between quantum phenomena with thermodynamics is not restricted to physics in curved spacetimes, in particular BH metrics, showing instead that even in flat space,
a uniformly accelerated observer would perceive the Minkowski vacuum as a thermal bath at temperature:
\begin{equation}
 T = \frac{\hbar a}{2\pi ck},
\end{equation}
where $a$ represents the acceleration of the observer. \par
Detector models have proven to be an invaluable and proefficient tool for understanding quantum phenomena in gravitational backgrounds. Numerous authors have investigated the response of various detector models in 
different scenarios, both in Minkowski space \cite{a7,a8,a9,a10} and spacetimes with curvature \cite{a14,a11,a12}. A third category of scenarios is represented by spacetimes with nontrivial topology \cite{a0,a2,a13}.
In the latter	 case the non-zero response of the detector has a component which is due to neither the motion, nor gravity, but has to do rather with the specific boundary conditions imposed by the topology of the spacetime. \par
In this thesis we present an analysis of the reponse of such a detector in a (mostly) flat spacetime which contains a topological feature called a ``wormhole''. \par
In the first chapter we briefly review the basics of quantum field theory on curved spacetime, highlighting some of the problems that can arise. Next we expose the basic elements of the theory of particle detectors, 
with emphasis on the Unruh-deWitt particle detector model, which we shall use. We present the transition probability and transition rate for the general case, for an arbitrary trajectory. \par
In the second chapter we review some of the results from the literature regarding the response of detectors in Minkowski space under various conditions. The transition rate of eternally static and eternally uniformly accelerating
detectors is well known. In more realistic scenarios, however different additional complexity is introduced. One class of nontrivial situations is represented by trajectories with varying velocities and acceleration, variations
which can be either smooth or sudden. Another scenario is when the interaction is switched on i.e. (the detector is ``coupled'') at a finite time, as opposed to being coupled in the infinite past. Both sets of ingredients act 
as perturbations for the quantum field and hence will produce spurious temporary variations in the response of the detector. \par
In the third chapter we present our results for the case of a spacetime containing a topological feature, called a ``wormhole''. We investigate the response of the detector on various trajectories, of which the most intriguing case is
when the detector crosses the throat of the wormhole. We use the Unruh-deWitt detector model in interaction with a massless scalar field. We find the appropriate solutions of the Klein-Gordon equation in the WH spacetime, 
in terms of spherical modes. Due to the nontrivial topology introduced by the WH, the transition probability of the intertial detector has strong variations at the moment it crosses the wormhole, followed by a period with tranzitory
oscillations, finaly stabilizing at a constant, nonzero value. 

\chapter{Modeling a particle detector}

The outcome of any physical calculation should be a measurable quantity. In quantum field theory these are usually represented by transition probabilities of scattering processes (with scattering understood with the general meaning of 
interaction). Another way of obtaining measurable quantities from quantum fields is by taking a device and interacting it with the quantum field and measuring the reponse of the device. One category of such useful theoretical tools are called
particle detectors. Basically these represent simple quantum mechanical systems which can make transitions between energy levels during their interaction with a quantum field. For example, if the field is in its ground state while it interacts 
with the detector, the response of the latter should be null. Conversely, if the field is in an excited state, the detector can engulf a quanta of it and transition to another energy. This is very useful in the context of general relativity,
where no universal notion of particle can be defined in general, in opposition to the situation on flat-space. The simplest case of such a detector is a quantum system (particle in a box), with two energy levels. The relevant quantity 
is the probability (and rate) of transition between the energy levels, which marks the presence (or absence) of excitations in the quantum fields to which it is coupled.  

\section{The quantization of fields on curved spacetimes}
\indent In this section we give the basic ingredients for quantizing fields, through the canonical procedure, on arbitrary curved manifolds and illustrate some of the difficulties one encounters. 
We use the general prescription for (interacting) quantum fields presented in the classical textbook by N.Birrell and P.Davis \cite{a1}. The book also contains the basic theory of particle detectors
along with a discussion about the ambiguities one faces in defining what is meant by ``particle'' in curved spacetime and how this can be circumvented by using detectors.    

Formally the process of canonical quantization of fields on an arbitrary spacetime is very similar to the case in Minkowski space. The first step is writing down the Lagrangian. For simplicity and use in this thesis we consider 
only the case of a scalar field. The basic substitutions necessary to pass from flat Cartesian to an arbitrary geometry are: 
\begin{eqnarray}
\nonumber \eta_{\mu\nu}&\rightarrow \ \ g_{\mu\nu}  \\
\partial_\mu &\rightarrow \ \ \nabla_\mu \ \,
\end{eqnarray}
where  $g_{\mu\nu}$ represents the metric tensor, $\eta_{\mu\nu}$ the Minkowski metric, and $\nabla_\mu$ the covariant derivative associated with the metric.

We denote with $m$ the mass of the scalar field, with $\xi$ the coupling of the field to gravity, while $R$ represent the Ricci constant. Keeping in mind that for a scalar field the covariant derivative is equal to the 
partial one, the Lagrangian of such a system is written as:
\begin{equation}
\mathcal{L} = \frac{1}{2\sqrt{-g}}[g^{\mu\nu}\varphi_{,\mu} \varphi_{,\nu}-(m^2+\xi R)\varphi^2].
\end{equation}
Using the Lagrangian, we can write down the action, the anulling of which with respect to variations of the field $\varphi$ leads to the field equations: 
\begin{equation} \label{eq:scalar_field_eq}
(\square+m^2+\xi R)\varphi = 0,
\end{equation}
where the d'Alembertian written in the metric $g_{\mu\nu}$ has the form:
\begin{equation}
\square = \frac{1}{\sqrt{-g}}\partial_\mu(\sqrt{-g} g^{\mu\nu}\partial_\nu)
\end{equation}
The $\xi R\varphi^2$ term represents the coupling between the field and the metric. The most widely used value for the coupling is $\xi = 0$, which represents the minimal coupling. Another popular choice is $\xi = 1/6$ which is called 
the conformal coupling because if the field is massless, the action and field equations are invariant under a conformal transformation of the metric (accompanied by a rescaling of the field):  
\begin{eqnarray}
\varphi &\rightarrow& \Omega^{-1}(x)\varphi \nonumber \\
g_{\mu\nu} &\rightarrow & \Omega^2(x) g_{\mu\nu},
\end{eqnarray}
where $\Omega$ is a real valued funuction of the coordinates, and is continuous and non-singular. \par

We consider heceforth a massless, minimally coupled scalar field. The Klein-Gordon equation reduces to the box operator: $ \square \varphi = 0$. \par
Assume that there exists a complete set of solutions $u_i$ of the above equations, that obeys the standard commutation relations, on constant spatial hypersurfaces $\Sigma$. The index ``i'' symbolizes the complete set of labels
necessary to unequivocally defined the functions $u_i$. Any general solution $\varphi$ of the field equation \eqref{eq:scalar_field_eq}, can be written in terms of the basis of functions $u_i$ as an expansion:
\begin{equation}
\varphi(x) = \sum_{i} \left(a_i u_i(x) + a^\dag_i u^*_i(x)\right)
\end{equation}
Applying the the canonical quantization procedure, the coefficients $a^\dag$ and $a$ become creation and annihilation operators, and they will be subjected to the commutation relations: 
\begin{equation}
[a_i,a_j^\dag] = \delta_{ij},\qquad  [a_i,a_j]=[a^\dag_i,a^\dag_j]=0.
\end{equation}
With the help of the operators $a$ and $a^\dag$ we can construct the entire Fock space. In particular, the vacuum state is defined as the state annulled by the 1-particle annihilation operator:    
\begin{equation}
 a_i\vert 0\rangle = 0
\end{equation}
\par
At this point we encounter the first difficulties. These are related to the fact that the set of mode functions $u_i$ are not unique. On the contrary, there are an infinity of sets of modes that obey the field equations. Consider
for example another set $\tilde{u}_i$. The field operator can be expanded identically in terms of this basis as follows: 
\begin{equation}
\varphi(x) = \sum_{i} \left(\tilde a_i \tilde u_i(x) +\tilde a_i^\dag\tilde u^*_i(x)\right)
\end{equation}
Given that both sets are complete, we can write one base as a function of the other as:
\begin{eqnarray}
\tilde u_j &= \sum\limits_{i}\ (\alpha_{ji} u_i + \beta_{ji} u^*_i) \nonumber \\
u_j &= \sum\limits_{i}\ (\alpha^*_{ji}\tilde u_i - \beta_{ji}\tilde u^*_i)
\end{eqnarray}
The above relations are called Bogolubov transformations, and the coefficients connecting the two bases are usually callled Bogolubov coefficients. 
From the above relations, and using the ortonormality of the modes, we find the relations between the two sets of annihilation and creation operators:
\begin{eqnarray} \label{eq:bogoliubov_transf}
a_j &= \sum\limits_{i}\ (\alpha^*_{ji}\tilde a_i + \beta_{ji}\tilde a^\dag_i) \nonumber \\
a_j^\dag &= \sum\limits_{i}\ (\alpha_{ji} \tilde{a}^\dag_i - \beta_{ji}^* a_i)
\end{eqnarray}
The coefficients $\alpha_{ji}$ and $\beta_{ji}$ are (complex) c-numbers. 
From eq.~\eqref{eq:bogoliubov_transf} we can see that in general the Fock spaces defined by the two sets of modes are not equivalent. In particular, if we look at the vacuum state of one set:  
\begin{eqnarray}
\tilde a_j\vert\tilde0\rangle = 0 \ \ \ \ \ \ \ \ \ \ \ \ \ \ \ \ \, \\
a_j\vert\tilde0\rangle = \sum\limits_{i}\beta^*_{ji}\vert\tilde1_j\rangle\neq0
\end{eqnarray}
we see that it will only coincide with the vacuum of the other set if $\beta_{ji} = 0$. In order to understand the physical interpretation of this result, we look at the average value of the particle number operator $N_i=a_i^\dag a_i$ of 
the mode $u_i$, in the vacuum state  $\vert\tilde0\rangle$ defined by the set of modes $\tilde{u}_i$:
\begin{equation}
\langle\tilde0\vert N_i\vert\tilde0\rangle = \sum_{j}\vert\beta_{ij}\vert^2.
\end{equation}
From the above relation we understand that the vacuum state $\vert \tilde{0} \rangle$ contains $\sum_{j}\vert\beta_{ij}\vert^2$ excitations of the  modes $u_i$.  
In the case of Minkowski space the problem is superficial, because a preferred coordinate system can be defined, to which we can associate a ``natural'' set of modes. This system is the Cartesian one $\{t,x,y,z\}$, selected
by the Poincar\' e group, the line element being invariant under the action of this group. The vector $\partial_t$ is a Killing vector of the underlying spacetime, orthogonal to the spatial hypersurfaces of constant time $t = const.$,
and the modes are eingenfunctions of this vector with the eigenvalue $-i\omega$ (for $\ \omega>0$, these are positive frequency modes). The physical content of this statement is the following: the existence of a preferred set of solutions 
allows us to uniquely define modes of positive/negative frequency, the excitations of which we will interpret as particles/antiparticles. The fact that the Killing vector $\partial_t$ is time-like on the whole manifold and the modes being
eingfunctions of this operator, guarantees us that the separation of positive/negative frequencies will not be affected during the time evolution. In other words, the modes defining particles and anti-particles will not get mixed during the
evolution. The vacuum in this case is invariant under Poincar\' e transformations, in other words it is the vacuum seen by all inertial observers. \par
In an arbitrary curved spacetime, the situation is much more problematic. The Poincar\' e group is no longer a symmetry group of the spacetime and in general there will not exist a time-like Killing vector with the help of which to define
positive frequencies. In the case of some spacetimes we can have a number of symmetries, like invariance to translations or rotations, in which case we can define natural coordinates associated to the Killing vectors, which are analogous 
the the rectangular coordinates of Minkowski space, but these will not have the same priviledged statute like we have in flat space. This is the case in general even for maximally symmetric spacetimes like deSitter and anti-deSitter space. 
 
 \section{The theory of the Unruh-deWitt detector}
 One of the most pressing issues in quantum field theory on curved spacetime is how to define clearly the notion of particles. As we have seen, in Minkowski space they are defined with the help of the symmetry group, the Poincar\' e group.
Thus, a mode of the field equation which describes a physical particle, is a positive frequency mode with respect to the usual time coordinate $t$. Under the action of the Poincar\' e transformations the positive frequency solutions always 
tranform into positive frequency solutions, thus all inertial observers can agree on what a particle means in the sense that all such observers measure the same number of particles in a given situation. Furthermore, the vacuum state, as 
defined by these modes is invariant under Poincar\' e transformations. The problems arising in an arbitrary curved spacetime become obvious. General relativity tells us that all coordinate systems are equivalent, and thus there is in general
no preferred time coordinate with respect to which we could define something as being of positive frequency. Even worse, the Poincar\' e group is no longer a symmetry group in an arbitrary spacetime, and there is no way of mapping onto it 
the concepts from flat space. \par
A possible solution, sugested by Unruh, de Witt, and others, was to think of particles in a pragmatic operational way, by defining a particle as being something which is ``detected'' by a ``particle detector''.  \par
The detector originally considered by Unruh (1976) represents a particle in a box coupled to a quantum field in a curved background. We say that a particle has beed detected when the detector, under the influence of the quantum field, 
makes a transition between energy levels corresponding to the absorption or emission of a quanta of the field. A very popular variant was introduced by de Witt (1979), which is a simpler version of Unruh's original idea.
The latter consists of a point-like quantum particle (usually having two energy levels) which is coupled to the quantum field via a monopole-type operator. In what follows we shall present briefly the basics of the theory of the Unruh-deWitt
particle detector. \par
The two main ingredients are: i) a scalar field $\varphi(x)$ and ii) the detector, which is a quantum system with two energy-levels $ \{\vert E_0\rangle,\vert E\rangle\}$  
 
Without lack of generality, we can consider the ground state energy of the detector as being $E = 0$, and we will denote this state with $\vert 0_{d}\rangle$.
The trajectory of the detector is $\{ t(\tau), \bf{x}(\tau) \} $, and we consider the interaction between the detector and the field as being point-like (characteristic of monopole interaction) and localized along the trajectory of the detector.
The Hamiltonian which describes the interaction has the following form:
 \begin{equation} \label{eq:int_hamiltonian}
 H_{int} = gm(\tau)\varphi(\tau),
 \end{equation}
 where $g$ is a coupling constant and $m(\tau)$ is the monopole moment of the detector. The evolution of the monopole operator is governed by the the Hamiltonian of the detector $H_{d}$:
  \begin{equation}
 m(\tau) = e^{iH_{d}\tau}m(0)\,e^{-iH_{d}\tau}
 \end{equation}
 Before we continue, we add a note on the nature of the interaction. Formally the interaction Hamiltonian \eqref{eq:int_hamiltonian} is written as:
 \begin{eqnarray} \label{eq:Hint}
  H_{int}(\tau) = g\int d^4x\, \rho(x)\varphi(x). 
 \end{eqnarray}
In the case of the monopole detector, the density operator is:
\begin{equation}
\rho(x) = m(x)\delta^{(4)}(x^\mu - x^\mu(\tau)),
\end{equation}
and the interaction Hamiltonian reduces to \eqref{eq:int_hamiltonian}. Note that the Dirac delta function enforces the point-like nature of the interaction on the trajectory of the detector $x^\mu(\tau)$.
The point-like nature is of course an idealization, all physical systems being of non-vanishing spatial extension. This can be taken into account by swapping the delta function for a smearing function 
or any other function which describes the spatial extent of the detector. We will take this insight into consideration in the following section. We continue for now with the monopole detector. \par
We consider the system formed by the detector along with the scalar field in the state $\vert 0_{d}\rangle\vert 0_\varphi\rangle = \vert 0\rangle$, where we have denoted with $\vert 0_\varphi\rangle$ the vacuum state
of the field $\varphi$, at time $\tau_0$. The quantity we are interested in is the probability that the detector will be found in the excited state $\vert E \rangle$ (different from the ground state $\vert E_0\rangle = \vert 0_{d}\rangle$) 
at a given later time $\tau_1 > \tau_0$, no matter what the state $\vert \varphi \rangle$ of the field is. In order to obtain this probability, we work in the interaction picture where all operators are evolved with the free Hamiltonian,
while the states are evolved using the interaction Hamiltonian. The transition amplitude of interest is given by:
 \begin{equation}
 A_{\, vac. \rightarrow\, \varphi,E} = \langle \varphi,E\vert \hat T\ exp\left[-i\int_{\tau_0}^{\tau_1}H_{int}(\tau)\, d\tau \right]\vert 0_\varphi,0_{d}\rangle
 \end{equation}
 where $\hat T$ is the time-ordering operator, and the transition we are considering is from the initial state $\vert 0_\varphi,0_{det}\rangle$ at time $\tau=\tau_0$, to the final state $\vert \varphi,E \rangle$ at time $\tau=\tau_1$.
 We can used perturbation theory to with the usual expansion of the time evolution operator into an inifinite sum of time-ordered products. The coupling constant being in general very small, our main interest is in the first order 
 term, which takes the form: 
 \begin{equation}
 A_{\, vac. \rightarrow\, \varphi,E} = ig\langle \varphi,E\vert \int_{\tau_0}^{\tau_1} d\tau\, m(\tau) \varphi(\tau) \vert 0_\varphi,0_{d}\rangle
 \end{equation}
 \begin{equation}
 \qquad \qquad \qquad \quad\ =ig\langle E\vert m(0)\vert 0_d\rangle \int_{\tau_0}^{\tau_1} d\tau\, e^{iE\tau} \langle\varphi\vert\varphi(\tau)\vert0_\varphi\rangle
 \end{equation}
 In order to find the probability for the detector to perform a transition, indifferent to what the final state of the field is, we must take the sqared modulus of the amplitude and sum over all possible final states $\vert\varphi\rangle$
 of the field. 
 \begin{equation}
 \sum_{\varphi}\vert\langle \varphi,E\vert0_\varphi,0_d\rangle\vert^2 = g^2\vert\langle E\vert m(0) \vert 0_d\rangle\vert^2 \int_{\tau_0}^{\tau_1}d\tau \int_{\tau_0}^{\tau_1} d\tau'\, e^{-iE(\tau-\tau')}\, \langle 0_\varphi\vert \varphi(\tau)\varphi(\tau') \vert 0_\varphi\rangle
 \end{equation}
 The factor preceding the integrals, called by some authors the ``selectivity'' of the detector, depends on the internal details of the detector and is hence independent of the trajectory of the detector and is thus of no physical interest to us
 here. The second term is called the response function of the detector and is universal in the sense that it does not depend on the details of the detector. 	 
 \begin{equation} \label{eq:transition_prob}
 \mathcal{F}_{\tau_0,\tau_1}(E) = \int_{\tau_0}^{\tau_1} d\tau \int_{\tau_0}^{\tau_1} e^{-iE(\tau-\tau')} d\tau' \langle 0\vert\varphi(\tau)\varphi(\tau')\vert 0\rangle
 \end{equation}
 Next, we follow the steps taken in  Schlicht \cite{a3}, and transform to the following set of integration variables: $u=\tau, s=\tau-\tau'$ for $\tau<\tau'$ and $u=\tau',  s=\tau'-\tau$ for $\tau'>\tau$. Thus the response function becomes:
 \begin{equation}
 \mathcal{F}_{\tau_0,\tau_1}(E) = 2 \int_{\tau_0}^{\tau_1}du \int_{0}^{u-\tau_0} ds\,\Re\left(e^{-iEs} \langle 0\vert \varphi(u)\varphi(u-s)\vert 0\rangle\right),
 \end{equation}
 where we have used the fact that $\langle 0\vert \varphi(\tau)\varphi(\tau')\vert 0 \rangle = \langle 0\vert \varphi(\tau)\varphi(\tau')\vert 0 \rangle^*$, and $\Re$ denotes the real part of the encased expression. Relabeling $\tau_1 = \tau$
 and differentiating with respect to $\tau$,  we find the following expression for the transition rate of the detector:
 \begin{equation}
 \dot{\mathcal{F}}_{\tau_0,\tau}(E) = 2 \int_{0}^{\tau-\tau_0}ds\, \Re\left(e^{-iEs} \langle 0\vert \varphi(\tau)\varphi(\tau-s)\vert 0\rangle\right)
 \end{equation} 
 Note that the rate has contributions only in the interval $[\tau_0,\tau]$, i.e. only the past evolution of the detector contributes and in this sense it is causal. If the function  $\langle 0\vert \varphi(\tau)\varphi(\tau')\vert 0\rangle$ is
 invariant under translation in time, in other words, if it depends only on $\Delta\tau = \tau-\tau'$, the expression of the transition rate can be simplified to the form \cite{a0}:
 \begin{equation} \label{eq:transition_rate}
 \dot{\mathcal{F}}_{\tau_0,\tau}(E) = \int\limits_{-\Delta\tau}^{\,\Delta\tau}ds\, e^{-iEs}\, \langle 0\vert \varphi(s)\varphi(0)\vert 0\rangle,
 \end{equation}
 where we have now denoted $\Delta\tau = \tau -\tau_0$. \par
 The transition rate is the only meaningful measurable quantity in this setting. In order to understand the physical meaning of the rate of transition, we write it in the following form:
 \begin{equation} \label{eq:physical_rate}
 \dot{\mathcal{F}}_{\tau_0,\tau}(E) = \lim_{\delta\tau \to 0} \frac{\mathcal{F}_{\tau+\delta\tau,\tau_0}-\mathcal{F}_{\tau,\tau_0}}{\delta\tau}
 \end{equation}
 From the above formula it becomes evident that the transition rate compares the response of detector from two coherent ensambles of detectores, one set of measurements being taken at $\tau$, while the other at $\tau + \delta\tau$.
 The response function $\mathcal{F}$, being proportional to a probability, is strictly non-negative. The transition rate, on the other hand can have domains on which it takes negative values (i.e. the probability decreases), this being 
 a hallmark of quantum phenomena. The strict condition that the rate has to obey is that its integrated value over the whole trajectory (i.e. the probability) has to be non-negative. 
 
\chapter{The response of the detector in Minkowski space}
As a preamble, in what follows, we use the Unruh-deWitt particle detector model in interaction with a massless scalar field, and investigate the response of a detector on different trajectories on Minkowski space. 
The transition rate has been thoroughly investigated in the literature, for inertial as well as accelerated trajectories. When one looks only at the transition rate, however there are some aspects related to the point-like nature 
of the detector which are glossed over. We perform a brief discussion of these issues. 
\section{The regularization of the Wightman function}
The scalar field is a solution to the massless Klein-Gordon equation.We can expand the field operator in terms of a standard set of ortonormal solutions of the field equation:
\begin{equation} \label{eq:mode_expansion}
\varphi(t,{\bf x}) = \int \frac{\ d^3k}{(2\pi)^\frac{3}{2}\sqrt{2\,\omega}}\, \left(a({\bf k}) e^{-i(\omega t + {\bf kx})} +a^\dag({\bf k}) e^{i(\omega t + {\bf kx})}\right),
\end{equation}
where for a massless field we have $\omega = \vert {\bf k}\vert$. \par
For the calculation of the transition rate \eqref{eq:transition_rate} we need to find the expression of the correlation function $ \langle 0\vert\varphi(x)\varphi(x')\vert 0\rangle$, which can be obtain using the mode expansion \eqref{eq:mode_expansion}
of the field. The correlation function, often called the Wightman function in the literature, takes the form:
\begin{eqnarray} \label{eq:wightman}
G(x,x') &\equiv&  \langle 0\vert\varphi(x)\varphi(x')\vert 0\rangle \\
&=& \frac{1}{(2\pi)^3} \int \frac{d^3k}{2\,\omega}\ \left( e^{-i\omega(t-t')+i{\bf k(x-x')}}\right)
\end{eqnarray}
The integral in \eqref{eq:wightman} can be solved by passing to spherical coordinates. The radial integral (with respect to $\vert \bf{k} \vert$) has an ultra-violet (UV) divergence which has to be eliminated by regularizing the integral.  
The usual regularization, known in the literature as the $\varepsilon$ prescription, consists in an exponential cut-off for the the high frequency modes, realized by introducing the term  $e^{-i\varepsilon \vert{\bf k}\vert}$ under the integral in
\eqref{eq:wightman}. After performing the integrals, we arrive at the following expression:
\begin{equation} \label{eq:wightman_ie}
G(x,x') = -\frac{1}{4\pi^2} \frac{1}{\left[(t-t'-i\varepsilon)^2-({\bf x-x'})^2\right]},
\end{equation}
where it is understood that $t=t(\tau), {\bf x}={\bf x}(\tau)$ and $t'=t'(\tau'), {\bf x}'={\bf x}'(\tau')$. \par
Schlicht (2004) has shown that the correlation function \eqref{eq:wightman_ie} is necessarily incorrect. The main argument is that employing the correlation function in this form, for a detector on a uniformly accelerated trajectory,
with the detector coupled (i.e. ''turned on'') at $\tau_0= -\infty$ the transition rate results in an expression dependent on the measurement time $\tau$, instead of taking the well known (thermal) time-independent form (to be described below).
The proposed solution consists in modifying the correlation function by employing a different regularization method. The new regularization is found by considering the point-like detector as arising in the limit of vanishing spatial extension. 
A powerful argument for this new method is that our theoretical detector model should be as close to reality as possible (while at the same time remaining as simpel a model as possible for it to allow an analytical approach). Intuitively,
the spatially extended detector should give a more physically sound result than the point-like model.  \par
The assummption underlying Schilcht's approach is the consideration of the detector as being a spatially extended object, which remains rigid in the detector's proper frame. The correct approach for building the proper frame, when considering
trajectories which can have non-vanishing proper-acceleration, is the employment of the Fermi coordinates  \cite{a15}. The immediate consequence of the presence of a minimal length-scale for the detector is the fact that modes with 
wavelength smaller than this scale can not be probed. Mathematically this can be mimicked by replacing the field operator defined at the point $x(\tau)$ in the interaction Hamiltonian with its value averaged over a characteristic volume
$V_\varepsilon$, centered on the point: 
 \begin{equation}
 \Phi(\tau) = \frac{1}{V_\varepsilon}\int_{V_\varepsilon} d^3\xi\, \varphi(x(\tau,\xi)),
 \end{equation}
 where $(\tau, \xi)$ represent the Fermi coordinates. The effect of the spatial averaging is the introduction of a cut-off at small distances, which translate into an incapacity to detect modes of high frequencies. Replacing the field
 field with $\Phi(\tau)$ in \eqref{eq:wightman}, we can now calculate the new regularized Wightman function:
 \begin{equation}
\langle 0\vert\Phi(x)\Phi(x')\vert 0\rangle = \frac{1}{(2\pi)^3} \int_{V_\varepsilon} d^3\xi\, d^3\xi' \int \frac{d^3k}{2\,\omega}\ \left( e^{-i\omega(t-t')+i{\bf k(x-x')}}\right),
\end{equation}
where it is understood that $x = x(\tau, \xi)$. Notice that in the limit $\varepsilon \rightarrow 0$ this expression reduces to the previous form \eqref{eq:wightman}.  
The elementary averaging introduced above can be generalized to a weighted average by employing a window function with characteristic length-scale $\varepsilon$:
\begin{equation}
 \Phi(\tau) = \int d^3\xi\, W_\varepsilon(\xi)\, \phi(x(\tau,\xi))
 \end{equation}
 
 We can use a window function with an infinite support, if the function falls off fast enough at large distances, or we can consider a truly finite dimension for the detector by using a function which is strictly non-zero only on a finite 
 domain. Either way, the window function has to be normalized:
 \begin{equation}
 \int d^3\xi\, W_\varepsilon(\xi) = 1,
 \end{equation}
 Note that the elementary averaging introduced previously is obtained if $W_\varepsilon = 1/V_\varepsilon$. In order for the weighted field operator to reduce to the non-averaged one, the window function has 
 to reduce to a Dirac delta function in the limit $\varepsilon \rightarrow 0$. The window function can be chosen in various ways, but the one considered by Schlicht is advantageous because the integrals 
 in the expression of the correlation function can be computed exactly. This function has the form:
 \begin{equation}
 W_\varepsilon(\xi) = \frac{2}{\pi^2} \frac{\varepsilon}{(\xi^2+\varepsilon^2)^2},
 \end{equation}
 sometimes called the Lorentz window function. The regularization induced by it is similar to the regular $i\varepsilon$ prescription, with the important difference that in this case the cut-off is performed in the proper frame of the detector,
 while the usual cut-off is done in an inertial ferefence frame. For inertial motion the two regularizations are equivalent. On the other hand, when accelerated trajectories are considered, significant differences arises. After some extensive 
 algebraic calculations we obtain the improved correlation function:
 \begin{equation} \label{eq:wightman_schlicht}
 G(x,x') = -\frac{1}{4\pi^2} \frac{1}{(t-t'-i\varepsilon(\dot{t}+\dot{t}'))^2-({\bf x-x'}-i\varepsilon(\dot{\bf x}+\dot{\bf x}'))^2},
 \end{equation} 
 where $\{t,{\bf x}\}$ and $\{t',{\bf x'}\}$ are functions of $\tau$ and $\tau'$, respectively. \par
 Henceforth we work with the correlation function \eqref{eq:wightman_schlicht}.  

\section{Inertial trajectories}
We begin by considering the response of the detector on inertial trajectories. We obtain the transition rate \eqref{eq:transition_rate} by making use of the correlation function \eqref{eq:wightman_schlicht}.
The $\varepsilon$ parameter in the correlation function, being the characteristic length of the detector, we shall fix to a small finite value. The investigation of the vanishing limit $\varepsilon \rightarrow 0$ being relegated to the end of the 
section.
\paragraph{\emph{Static detector}}
\ \\
\\
For a detector at rest the trajectory is $t = \tau ,\, {\bf x} = {\bf x}_0 = const $. We consider the detector as ``functioning'' at all times, i.e. it is coupled at $\tau_0 = -\infty$. In such conditions the transition rate is:
\begin{equation}
\dot{\mathcal{F}}_\tau (E) = -\frac{1}{4\pi^2} \int\limits_{-\infty}^{\,\infty}ds\,\frac{e^{-iEs}}{(s-2i\varepsilon)^2}.
\end{equation}
As mentioned earlier, the correlation function in this case is equivalent with the one obtained through the classical $i\varepsilon$ regularization.
The integral can be evaluated using the theorem of residues. The function has a pole on the positive imaginary semi-axis, and thus for $E > 0$ the rate is zero. The case $E < 0$ would correspond to a transition to an inferior energy level
of the detector, in other words it would signal a spontaneous emission. We have ruled out this possibility by considering the detector with only two levels, with the detector initially being in the ground state $\vert 0_d\rangle$. 
The result is intuitively-obvious: a detector at rest for eternity in empty flat space will not detect anything. \par
We have considered the case of a detector at rest, but the result generalizes to all inertial trajectories. This is guaranteed by the fact that the generators of boosts are Killing vectors of Minkowski spacetime and thus 
all physical quantities have to be independent of velocity.

\paragraph{\emph{Sudden take-off}} \ \\ \\
We continue by investigating the case of a detector on an inertial trajectory, which suffers a sudden jump in the velocity. The physical picture is the following: from the infinite past the detector is at rest, then at a certain moment it gets
a sudden jolt and continues on an inertial uniform trajectory with constant velocity $v$ up to the infinite future. Considering the jump in the velocity as occuring at $\tau = 0$, the trajectory of the detector is:
\begin{eqnarray}
t(\tau) = \theta(\tau) \gamma \tau + \theta(-\tau)\tau  \nonumber \\
x(\tau) = \theta(\tau) \gamma v\tau, \ \ \ \ \ \qquad
\end{eqnarray}
where $\theta$ is the Heaviside function, $\gamma$ is the Lorentz factor and time varies between $\tau \in\{-\infty, \infty\}$. \par
Although the trajectory is inertial on both intervals, a non-vanishing rate will appear at the moment of the jump. This is possible because the correlation function is no longer invariant under time-translations (i.e. Lorentz invariance is lost).
We can anticipate the form of the transition rate: 
\begin{enumerate}
 \item before the jolt the detector is static, resulting in a vanishing rate 
 \item around  $\tau=0$, the moment of the jump, the rate will increase suddenly
 \item in the infinite future $\tau \rightarrow \infty$, the rate should fall back to zero
\end{enumerate}

\begin{figure}
\centering
\begin{subfigure}{.5\textwidth}
  \centering
  \includegraphics[width=.85\linewidth]{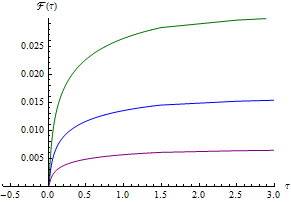}
  \label{fig:3.1a}
\end{subfigure}%
\begin{subfigure}{.5\textwidth}
  \centering
  \includegraphics[width=.95\linewidth]{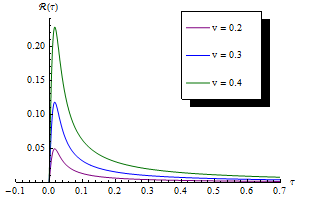}
  \label{fig:3.1b}
\end{subfigure}
\caption{\emph{Transition probability (left) and rate (right) with the detector evolving on a piecewise inertial trajectory with a sudden velocity jump, for $E = 0.1$ and the cut-off fixed at $\varepsilon = 0.01$.\ } }
\label{fig:3.1}
\end{figure}

\begin{figure}
\centering
\begin{subfigure}{.5\textwidth}
  \centering
  \includegraphics[width=.95\linewidth]{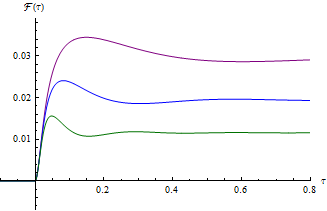}
  \label{fig:3.2a}
\end{subfigure}%
\begin{subfigure}{.5\textwidth}
  \centering
  \includegraphics[width=.95\linewidth]{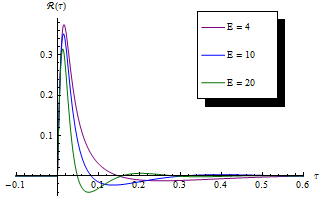}
  \label{fig:3.2b}
\end{subfigure}
\caption{\emph{The transition probabilitatea (left) and rate (dreapta) for the detector which is static up to the moment $\tau_0 =0$, point at which it starts moving with $v = 0.5$. The detector  length scale
is fixed at $\varepsilon  = 0.01$.\ }}
\label{fig:3.2}
\end{figure}

The sudden take-off should be interpreted as the limiting case of a sharp acceleration during a time interval, when this interval is taken to zero. The sudden change in the velocity acts as a perturbation for the field.
The upper limit for the frequency of modes which get excited during the process is linked to the time scale under which the acceleration takes place. Taking the vanishing limit equates to exciting field modes with infinitely 
large frequencies. On the other hand, we have employed in the Wightman function the regularization consistent with a spatially extended detector. Thus, the detector will be unable to absorb quanta which have associated length scales
smaller than the characteristic size of the detector. If we would consider the vanishing limit $\varepsilon \rightarrow 0$, such that the spatial extension no longer limits the absorption of arbitrarily high frequency modes, the probability
and rate diverges. We observe that it is the interplay between the properties of the trajectory and those of the detector which produces divergences. Indeed, the presence of a divergence  requires both: $a)$ a
sudden velocity variation in the trajectory, or the properties of the coupling, which has to be such that arbitrarily high frequency modes can be excited, and $b)$ a detector with the right properties, such that it can detect these modes. \par
We highlight one more aspect: notice that although the rate vanishes as we go to the infinite future as expected, the probability does not. Instead it tends to a constant value which increases as we decrease the spatial cut-off $\varepsilon$.
If we consider the transition rate as being the only physically measurable quantity and if the measurement is done at a large time after the velocity jump, the response is experimentally indistinguishable from that of the static detector.
Indeed the transition rate at $\tau \rightarrow \infty$ is zero in both cases. This suggests that one needs to look at both the rate and the probability of transition in such a setup in order to get the full picture. 

\paragraph{\emph{Finite-time trajectories}}
\ \\ \\
Until now we have considered that our detector is always ``functioning``, i.e. that the detector is coupled to the field at $\tau_0 \rightarrow -\infty$. In a more realistic scenario, we should look at detectors which are switched on
at a finite time $\tau_0$ and is operating on a finite interval $\Delta\tau$. In such a case, the response of the detector will be a combination of effects due to both the properties of the trajectory and those of the coupling. 
We consider an instantenous coupling, which can be modelled by considering the coupling constant as being time-dependent:
\begin{equation} \label{eq:coupling}
 g(\tau) = g_0\Theta(\tau-\tau_0)
\end{equation}
One might object that this breaks Lorentz invariance, but when we are talking about measurements and such we have already introduced a priviledged time coordinate.  \par
\begin{figure}
\centering
\begin{subfigure}{.5\textwidth}
  \centering
  \includegraphics[width=.95\linewidth]{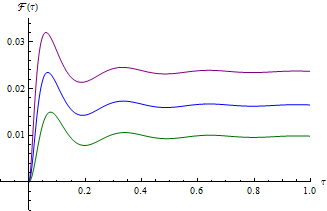}
   \label{fig:3.3a}.
\end{subfigure}%
\begin{subfigure}{.5\textwidth}
  \centering
  \includegraphics[width=.95\linewidth]{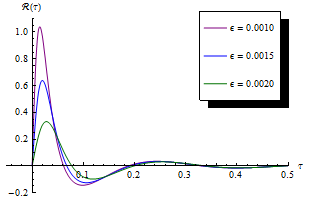}
  \label{fig:3.3b}h
\end{subfigure}
\caption{\emph{Transition probability (left) and rate (right) of the detector on a uniform trajectory, with the coupling switched on at $\tau_0=0$. \ \ $E = 1$, $ v =0.5$.} }
\label{fig:3.3}
\end{figure}

For simplicity we consider the detector as being static, and we are only switching it on and off. Using the coupling \eqref{eq:coupling} with $\tau_0 = 0$, we obtain the transition rate as:
\begin{equation}
\dot{\mathcal{F}}_{\tau_0,\tau} (E) = -\frac{1}{4\pi^2} \int\limits_{0}^{\,\Delta\tau}ds\,Re\left(\frac{e^{-iEs}}{(s-2i\varepsilon)^2}\right)
\end{equation}
We observe that the above expression ca be rewritten as a sum of two terms.
\begin{equation}
\dot{\mathcal{F}}_{\tau_0,\tau} (E) = -\frac{1}{4\pi^2} \left(\int\limits_{0}^{\,\infty}-\int\limits_{\Delta\tau}^{\,\infty} \right) ds\,Re\left(\frac{e^{-iEs}}{(s-2i\varepsilon)^2}\right)
\end{equation}
The first integral represents the contribution from the infinite proper time static trajectory, for which the respone of the detector is zero as shown above. The second term is a correction due to the coupling of the 
detector. Denoting the correction with $\dot{\mathcal{F}}_{\Delta \tau}$, we can write it as:

\begin{equation}
\dot{\mathcal{F}}_{\Delta\tau}(E) = -\frac{1}{2\pi^2}\int_{\Delta\tau}^{\infty} ds \frac{cos(Es)}{(s-2i\varepsilon)^2},
\end{equation}
which is equivalent with the result obtained by Svaiter and Svaiter \cite{a8}. The authors of Ref.~\cite{a8} note that this integral has domains on which it takes negative values. The important aspect is that the rate 
should vanish in the infinite time asymptotic limit and that the probability be positive on the whole trajectory, which can be seen from Fig.~\ref{fig:3.3}.

The response of the detector in the case of finite-time coupling significantly differs from the case of sudden take-off. In particular, in the case of finite-time coupling, the transition probability and rate are independent of the velocity
of the detector, the Wightman function being boost invariant. In this case we have considered a sudden coupling of the detector, described by the step function in \eqref{eq:coupling}. The situation is somewhat similar to the sudden take-off
scenario in that the coupling here takes place under an infinitely small time interval, with the consequence that infinitely large frequency modes get excited. If we would make the coupling function smoother by swapping the 
step function with an continuous function $\chi(\tau)$ (for example an $\arctan$), the upper limit of excitable modes would be of the inverse order of the time-scale of the coupling function. The probability and rate remain finite because we are working with a spatially extended detector
with the characteristic length scale given by the parameter $\varepsilon$. It can be seen from Fig.~\ref{fig:3.3} that as we decrease the length of the detector, the probability and rate increase. It is an interesting feature that in this case
the probability shows a larger variation with more pronounced spurious oscillations than in the case of sudden take-off Fig.~\ref{fig:3.2}, which also die off slower in this case. On the other hand, as was the case previously also, after
the oscillations are damped down, the transition rate vanishes while the probability tends to a finite constant value. In a physical setup the time of measurement should be taken at a time large enough for the spurious oscillations to vanish
and in the end the remaining constant probability should be subtracted from the result in order to separate the response due to the properties of the trajectory from those of the coupling.   
Note that if a smooth coupling is used, we can take the point-like limit of the detector without obtaining divergences, however the result will depend on the explicit form of the decoupling function $\chi(\tau)$. One then has to separate
the contribution pertaining to the coupling from that arising from the properties of the trajectory in order to obtain a meaningful physical result. The influence of the coupling function on the response of the detector was studied in Refs.~\cite{a9,a14,a16}.
The effects of different window functions for modeling the detector, within Schlicht's regularization method, was studied in \cite{a17}. \par

\pagebreak
\section{Accelerated trajectories}
\paragraph{\emph{Uniformly accelerated motion}}
\ \\ \\
In what follows, we consider a detector moving on a uniformly accelerated wordline, described by the trajectory:
\begin{eqnarray}
t(\tau) &=& \alpha\, \sinh\left(\,\tau/\alpha\, \right) \nonumber \\
x(\tau) &=& \alpha\, \cosh\left(\,\tau/\alpha\,\right),
\end{eqnarray}
where $\alpha$ is the 3-acceleration (which is constant), and the detector is coupled at $\tau_0 \rightarrow -\infty$. 
After a bit or reorganizing, we arrive at the following expression of the correlation function:
\begin{equation}
W(\tau, \tau') = -\frac{1}{(4\pi)^2} \frac{1}{\left(\alpha \sinh\left( \frac{\tau-\tau'}{2\alpha}\right) - i\varepsilon \cosh\left(\frac{\tau-\tau'}{2\alpha}\right)\right)^2}
\end{equation}
Exploiting the time-translation invariance of the correlation function, we can write the transition rate as:
\begin{equation}
\dot{\mathcal{F}}_\tau(E) = -\frac{1}{(4\pi)^2}\int\limits_{-\infty}^{\infty} ds\, \frac{e^{-iEs}}{\left(\alpha\sinh\left(\frac{s}{2\alpha}\right)-i\varepsilon\cosh\left(\frac{s}{2\alpha}\right)\right)^2}
\end{equation}
The integral can be solved by using the residue theorem, resulting in:
\begin{equation} \label{eq:thermal_rate}
\dot{\mathcal{F}}_\tau(E) = \frac{E}{2\pi}\frac{1}{e^{2\pi E\alpha}-1},
\end{equation}
which is the classical result well known in the literature. The rate respects the KMS condition:
\begin{equation}
\dot{\mathcal{F}}_\tau(E) = e^{-\frac{E}{T}} \dot{\mathcal{F}}_\tau(-E),
\end{equation}
where we have denoted $T = \frac{1}{2\pi\alpha}$. It is said that the responsed of the case of a (eternally) uniformly accelerating detector is thermal, because it is equivalent to the rate of a detector which is in a thermodynamic
equilibrium with a thermal bath (of particles) at temperature T. In this case there is the well known relation between spontaneous emission and absorption, reflected in the case of our accelerated detector by the KMS condition.

\paragraph{\emph{Sudden acceleration}}
\ \\ \\
The next case that we investigate is that of a detector which is static up to a moment of time $\tau_0$, point at which it starts accelerating with the constant acceleration $1/\alpha$. The trajectory is described by the relations:
\begin{eqnarray}
t(\tau) &=&  \alpha\, \sinh(\tau/\alpha)\, \Theta(\tau)+ \tau\,\Theta(-\tau) \nonumber \\
x(\tau) &=&  \alpha\, \cosh(\tau/\alpha)\, \Theta(\tau),
\end{eqnarray}
where $\Theta$ is the Heavisede (step) function, and we have considered that the detector starts accelerating at $\tau_0 = 0$. \par
\begin{figure}
\centering
\begin{subfigure}{.5\textwidth}
  \centering
  \includegraphics[width=.95\linewidth]{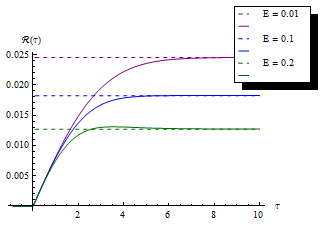}
  \label{fig:3.5a}
\end{subfigure}%
\begin{subfigure}{.5\textwidth}
  \centering
  \includegraphics[width=.95\linewidth]{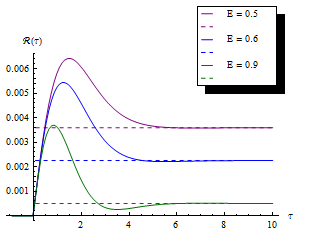}
   \label{fig:3.5b}
\end{subfigure}
\caption{\emph{ The transition rate (continuous lines) of the detector hat is static up to the moment $\tau = 0$, point at which it starts accelerating with uniform acceleration $\alpha = 1$. Superimposed are the thermal rates 
\eqref{eq:thermal_rate} for the eternally accelerated detector with the same acceleration (dotted lines). The detector length is fixed at $\epsilon = 0.01$.}}
\label{fig:3.5}
\end{figure}
The sudden acceleration acts as a perturbation, exciting the modes of the scalar field and producing transitory effects which die off as we go towards larger times. 
Note that as we increase the energy level of the detector the spurious oscillations become more prominent, as can be see in Fig.~\ref{fig:3.5} . In the asymptotic future limit the rates tend towards 
the corresponding thermal response of the eternally uniformly accelerating detectors with the same acceleration. 

\paragraph{\emph{Finite-time accelerated trajectory}} \ \\ \\
The last case that we look at is that of an eternally uniformly accelerated detector, wich is switched on at the moment $\tau_0$. We consider the coupling as happening instantaneously. The response of the detector can be separated into 
two distinct contributions, as is the case in the analogous inertial setup. The first of these is the response of the detector coupled on at $\tau \rightarrow -\infty$, while the second is a correction due to the finite time under
which the detector is ''detecting``.  
\begin{eqnarray} \label{eq:finite_acceleration}
\dot{\mathcal{F}}_{\tau_0,\tau} (E)  \nonumber \\
&=& -\frac{1}{(4\pi)^2} \left(\int\limits_{0}^{\,\infty}ds-\int\limits_{\Delta\tau}^{\,\infty}ds \right)\ \Re\left[\frac{e^{-iEs}}{\left(\alpha\,sinh\left(\frac{s}{2\alpha}\right)-i\varepsilon\,cosh\left(\frac{s}{2\alpha}\right)\right)^2}\right],
\end{eqnarray}
where the first integral is the thermal rate arising from the eternally uniformly accelerated motion \eqref{eq:thermal_rate}, while the second is the correction term, and is equal to:
\begin{equation}
\dot{\mathcal{F}}_{\Delta\tau}(E)= \frac{1}{(4\pi)^2} \int_{\Delta\tau}^{\infty} ds \frac{\cos(Es)}{\left(\alpha\sinh\left(\frac{s}{2\alpha}\right)-i\varepsilon\cosh\left(\frac{s}{2\alpha}\right)\right)^2}
\end{equation}

\begin{figure}
\centering
\begin{subfigure}{.5\textwidth}
  \centering
  \includegraphics[width=.95\linewidth]{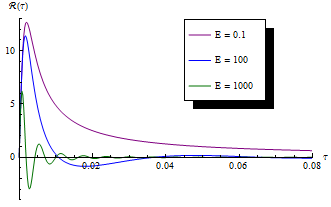}
  \label{fig:3.5a}
\end{subfigure}%
\begin{subfigure}{.5\textwidth}
  \centering
  \includegraphics[width=.95\linewidth]{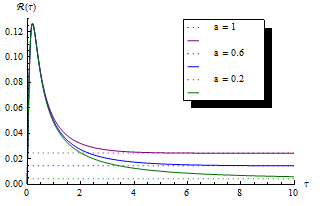}
   \label{fig:3.5b}
\end{subfigure}
\caption{\emph{ The transition rate (continuous lines) of the uniformly accelerating detector with acceleration $\alpha = 1$, that is switched on at $\tau = 0$. The sudden coupling produces spurious effects which 
die off in the asymptotic future limit, the rate reverting to the corresponding thermal expression \eqref{eq:thermal_rate} (dotted lines). The cut-off parameter  is $\epsilon = 0.01$. }}
\label{fig:3.6}
\end{figure}
The correction term, being an even function of $E$, does not respect the KMS conditions, and thus the detector following the trajectory \eqref{eq:finite_acceleration} is not thermal in this sense. In the asymototic limit  
$\tau \rightarrow \infty$ the correction term vanishes and thus the rate reverts to the thermal nature.

\section{The point-like limit}
\begin{figure}
\centering
\begin{subfigure}{.5\textwidth}
  \centering
  \includegraphics[width=.95\linewidth]{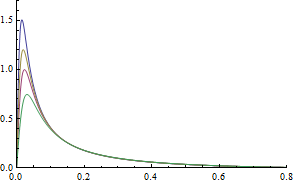}
\end{subfigure}%
\begin{subfigure}{.5\textwidth}
  \centering
  \includegraphics[width=.95\linewidth]{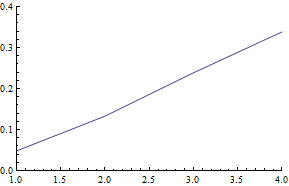}
\end{subfigure}
\caption{ \emph{ On the left panel the tranzition rate of a detector evolving on an inertial trajectory, with the coupling switched on at $\tau = 0$. The curves represent different values of the cut-off parameter $\eta$. 
The log-plot on the right shows the probability, at a sufficiently large future time such that the spurious effects induced by switch-on have dammed down. The probability is represented as a function of the inverse detector
length scale, confirming eq.\eqref{eq:IR_div}. }}
\label{fig:3.4}
\end{figure}

The consider now the vanishing limit of the regularization parameter $\varepsilon \rightarrow 0$, which in the paradigm used by Schlicht represents the characteristic length of the detector. In the vanishing limit the detector is 
point-like and can absorb arbitrarily high frequency modes (with arbitrarily small wavelength). If such modes are excited during the evolution along the trajectory, the transition probability for the detector will diverge. 
This can be seen from Fig.~\ref{fig:3.4}, where curves from different values of the detector length are shown, for the case of uniformly moving detector wich is switched on at the time $\tau_0=0$. We can see that the varying length scale
affects the transition rate only for a small time after the coupling is switched on, the rate falling off to zero as we increase the time. On the other hand, if we look at the probability at large times, this does not fall off but rather
tends to a constant value, which diverges as we take the point-like limit of the detector. \par 
 
To investigate the $\epsilon$-dependence of the detector response in the point-like limit, we take a few steps back and consider the probability of an eternally static detector. The correlation function in this case reduces to:
\begin{equation}
 G(x,x') \equiv - \frac{1}{4\pi^2}\frac{1}{(\tau -\tau' - i\varepsilon)^2},
\end{equation}
We begin by writing the probability for finite switch-on and off time, and take the infinite limit afterwards. In this case the probability \eqref{eq:transition_prob} is written as:
\begin{equation}
\mathcal{F}_{\tau_0,\tau_1}(E) = -\frac{1}{4\pi^2} \int_{\tau_0}^{\tau_1} d\tau \int_{\tau_0}^{\tau_1} d\tau'  \frac{e^{-iE(\tau-\tau')}}{(\tau -\tau' -i\varepsilon)^2}  
\end{equation}
We now follow Ref.~\cite{a8} and change to the variables $\tau^+ = \tau + \tau'$ and $\eta = \tau - \tau'$, and use the notaion $\Delta\tau = \tau_1 - \tau_0$. Notice that the integrand depends only on $\tau^-$, 
hence the $\tau^+$ integral can be done immediately. The probability thus becomes:
\begin{eqnarray}
 \mathcal{F}_{\tau_0,\tau_1}(E) &=& - \frac{1}{4\pi^2}\int_{-\Delta\tau}^{\Delta\tau} d\eta \left(\Delta\tau - \vert \eta \vert \right) \frac{e^{-iE\eta}}{(\eta - i\varepsilon)^2}  \\
 &\equiv& \mathcal{F}_{\Delta\tau}^{I} + \mathcal{F}_{\Delta\tau}^{II}
\end{eqnarray}
Next we take the limit of infinite proper time. The first integral can be evauluated by closing the contour in the complex plane, keeping in mind that $E > 0$. This results in an indeterminate expression when we take the limit,
on which we can use L'Hospital's theorem to obtain:
\begin{eqnarray}
 \mathcal{F}^{I} &\equiv&  \lim_{\Delta\tau \rightarrow \infty} \mathcal{F}_{\Delta\tau}^{I} \\
   &=& \lim_{\Delta\tau \rightarrow \infty}\Delta \tau \int_{-\Delta\tau}^{\Delta\tau} d\eta \frac{e^{-iE\eta}}{(\eta - i\varepsilon)^2} \\
   &=& \lim_{\Delta\tau \rightarrow \infty} -2\Delta\tau \int_{\Delta\tau}^{\infty} d\eta \frac{cos(E\eta)}{\eta^2} \\
   &\simeq& \lim_{\Delta\tau \rightarrow \infty} \frac{2\cos(E\Delta\tau)}{\Delta\tau^2} \\
   &=& 0
 \end{eqnarray}
 The second term is:
 \begin{eqnarray}
 \mathcal{F}^{II} &\equiv&  \lim_{\Delta\tau \rightarrow \infty} \mathcal{F}_{\Delta\tau}^{II} \\
   &=& \frac{1}{4\pi^2}\int_{-\infty}^{\infty} d\eta \, \frac{\vert\eta\vert e^{-iE\eta}}{(\eta - i\varepsilon)^2} \\
   &=& \frac{1}{4\pi^2}\left(\int_{0}^{\infty} - \int_{-\infty}^{0} \right) d\eta \, \frac{\eta e^{-iE\eta}}{(\eta - i\varepsilon)^2}
 \end{eqnarray}
Both integrals can be evaluated by integrating in the complex plane. Because $E > 0$, the contours have to be closed on the negative imaginary semi-axis and thus the integrand has no poles. Both integrals have the same form, such that we 
can write:
\begin{eqnarray} \label{eq:point_limit}
 \mathcal{F}^{II} &=& \frac{1}{2\pi^2} \int_{0}^{\infty}d\eta   \frac{\eta e^{-E\eta}}{(\eta + \varepsilon)^2}  \\
 &=& -\frac{1}{2\pi^2} \left(1 + e^{\varepsilon E}\left(\varepsilon E+1\right)\text{Ei}(-\varepsilon E)\right),
\end{eqnarray}
where the exponential integral function is defined as:
\begin{equation} \label{eq:exponential_integral}
 \text{Ei}(z) = -\int_{z}^{\infty} dt \frac{e^{-t}}{t}
\end{equation}

Note that the probability depends only on the combination $\varepsilon E$. The fact that this has to hold for any trajectory or setup can be argued already on the basis of units. \par
We can now take the limit of the probability for small $\varepsilon$ in expression \eqref{eq:point_limit} to find:
\begin{eqnarray} 
 \mathcal{F}(\varepsilon E \rightarrow 0) &=& -\frac{1}{2\pi^2}\left(1 + \gamma + \log(\varepsilon E)\right) \\
  &\simeq& -\frac{\log(\varepsilon E)}{2\pi^2}.  \label{eq:IR_div}
\end{eqnarray}
 Note that the above expression has no time dependence, and hence the rate of transition vanishes.

\chapter{The response of the detector in a thin-shell wormhole metric} 
In this chapter we investigate the response of the Unruh-deWitt detector in a spacetime that contains a topological structure, called a ''wormhole`` (WH). We look at different trajectories, the most interesting being one that 
radially crosses through the wormhole. To our knowledge this problem has not been investigated so far in the literature. 
\section{Wormhole geometry}
\begin{figure}
\centering
  \includegraphics[width=.95\linewidth]{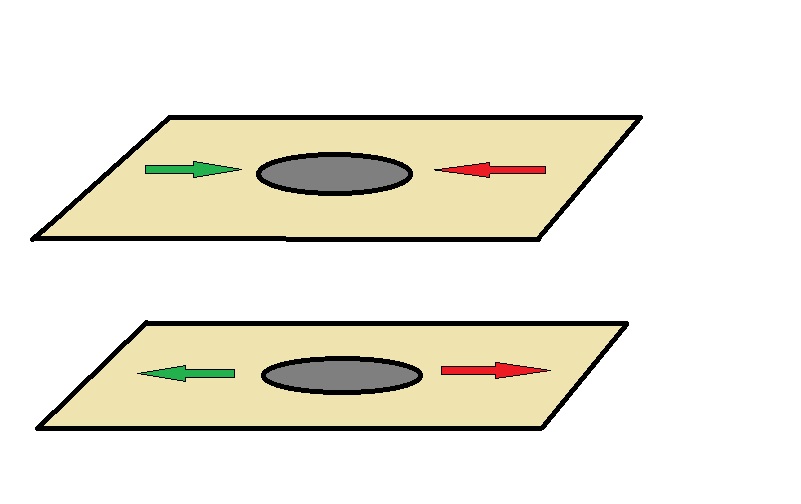}
\caption{\emph{ The geometry of the wormhole spacetime, described by the metric \eqref{eq:wh_metric}. The green and red arrows show the trajectory of linerly moving observers as they pass through the throat of the wormhole.}}
\label{fig:4.1}
\end{figure}

We consider the most elementary type of wormhole, known in the literature under the name ''thin-shell wormhole``, described by the metric:
\begin{equation} \label{eq:wh_metric}
ds^2 = -dt^2 + dr^2 + r^2(\rho)(d\theta^2 + \sin^2\theta d\phi^2),
\end{equation}
where the radial coordinate $r$ is:
\begin{equation} 
r(\rho) = a + \vert \rho \vert,\ \rho \in (-\infty,\infty)
\end{equation}
Note that the physical radial coordinate $\rho$ takes both positive and negative values. The metric \eqref{eq:wh_metric} describes two Minkowski spacetimes from which spheres of radius $a$ have been cut out, and with the two
manifolds being glued together along the surface of these spheres. In Fig.\eqref{fig:4.1} there is a schematic representation of this setup. The resulting spacetime is mostly flat, with the non-zero curvature being localized 
exlusively at $r = a$. The Ricci scalar is equal to:
\begin{equation}
R = -\frac{8}{a}\delta(\rho)
\end{equation}
\par

With the publication in (1988) of the paper \cite{a18,a19} by M.Morris and K.Thorne, the study of wormhole-related topics took off feverishly. Although at the present time there is no evidence of the existence of wormholes in 
the physical Universe, nevertheless their investigation can lead to important insights into the nature of gravity and the compatibility between quantum physics and general relativity. The properties of quantum fields in such 
geometries with localized curvature and interesting topological properties is perhaps the most interesting arena to probe. A significant part of the literature on wormholes is concerned with the stability of wormholes \cite{a20,a21}.
Alternative directions of research are: the propagation of waves in WH geometries \cite{a22}, self-force calculations in various scenarios \cite{a23} and the investigation of vacuum fluctuations in the presence of WHs \cite{a24}.
M.Visser's excellent book \cite{a25} is an invaluable resource for anyone interested in the topic. \par
At the present moment, to the best of our knowledge, there is no paper published on the subject of the Unruh effect in spacetimes containing wormholes. An extensive study of the response of particle detectors on various spacetimes
with nontrivial topology has been performed by P.Langlois \cite{a0,a2}. 
\section{Spherical modes}
We consider a massless scalar field, minimally coupled to gravity, obeying the Klein-Gordon equation:
\begin{equation} 
\left(\triangle-\frac{\partial^2}{\partial t^2}\right) \varphi({\bf x},t) = 0,
\end{equation}
where the d'Alembertian operator is written with the metric \eqref{eq:wh_metric}.\par
We apply a fourier transformation on the field $\varphi$:
\begin{equation}
\varphi({\bf x},t) = \int\limits_{-\infty}^{\infty} \varphi({\bf x},\omega) e^{-i\omega t},
\end{equation}
where each component is a solution of the Helmholtz-type equation:
\begin{equation} \label{eq:helmholtz}
(\nabla^2 + k^2)\,\varphi({\bf x},\omega) = 0.
\end{equation}
We have denote with $k^2 = \omega^2/c^2$. The problem has spherical symmetry, and thus it is appropriate to search for solutions corresponding to spherical coordinates. In such a coordinates system, the Laplace operator takes the form:
\begin{equation}
\nabla^2 = \frac{1}{r^2}\,\frac{\partial}{\partial r}\left( r^2\,\frac{\partial}{\partial r}\right) + \frac{1}{r^2 \sin\theta}\,\frac{\partial}{\partial\theta} \,\left(\sin\theta\,\frac{\partial}{\partial\theta}\right) + \frac{1}{r^2 \sin^2\theta}\,\frac{\partial^2}{\partial \phi^2}
\end{equation}
The solution to the scalar wave equation in terms of spherical waves can be found for example in Ref.\cite{a26}. By virtue of the spherical symmetry, we can separate the angular variables from the radial one. The general 
solutions of eq.\eqref{eq:helmholtz} take the form:
\begin{equation}
\varphi({\bf x},\omega) = \sum_{l,m} f_{lm}(r) Y_{lm}(\theta,\Phi),
\end{equation}
where $Y_{lm}$ represent the spherical harmonics, defined as:
\begin{equation} \label{eq:spherical_harmonics}
Y_{lm} = \sqrt{\frac{2l+1}{4\pi}\frac{(l-m)!}{(l+m)!}}\, P^m_l(cos\theta)e^{im\phi},
\end{equation}
while the functions $f_{lm}$ are solutions of the radial equation: 
\begin{equation} \label{eq:wh_radial_eq}
\left[\frac{d^2}{dr^2} + \frac{2}{r}\frac{d}{dr} + k^2 - \frac{l(l+1)}{r^2}\right] f_{lm}(r) = 0,
\end{equation}
with $r(\rho) = a+ \vert \rho\vert$. \par
In the case of non-minimal coupling, eq.\eqref{eq:wh_radial_eq} contains the additional term $\xi R=-8\xi\frac{\delta(\rho)}{a}$, due to the interaction with the gravitational field which is localized at $\rho = 0$. 
The radial equation can be solved piecewise on the two regions $sgn(\rho) = \pm$, where the equation is the usual radial equation from Minkowski space, and then connecting the solutions at the wormhole throat by imposing 
$\mathcal{C}^1$ continuity conditions. \par
Performing the substitution:
\begin{equation}
f_{lm} (r) = \frac{1}{\sqrt{r}} u_{lm}(r),
\end{equation}
equation \eqref{eq:wh_radial_eq} becomes:
\begin{equation}
\left[\frac{d^2}{dr^2}+ \frac{1}{r}\frac{d}{dr} + k^2 -\frac{(l+\frac{1}{2})^2}{r^2}\right]u_{lm}(r) = 0
\end{equation}
This represents a Bessel equation with $\nu = l+\frac{1}{2}$. Thus the general solutions to the radial equation \eqref{eq:wh_radial_eq} are: 
\begin{equation}
f_{lm} (r) = \frac{A_{lm}}{\sqrt{r}} J_{l+\frac{1}{2}}(kr) + \frac{B_{lm}}{\sqrt{r}}Y_{l+\frac{1}{2}}(kr)
\end{equation}
At this point it is natural to introduce the spherical Bessel and Hankel functions:
\begin{eqnarray}
j_{l}(z) &=& \sqrt{\frac{\pi}{2z}}\ J_{l+\frac{1}{2}} (z) \\
n_{l}(z) &=& \sqrt{\frac{\pi}{2z}}\ N_{l+\frac{1}{2}} (z) \\
h_{l}^{(1,2)}(z) &=& \sqrt{\frac{\pi}{2z}}\, [J_{l+\frac{1}{2}} (z) \pm i Y_{l+\frac{1}{2}} (z)]
\end{eqnarray}
\par
In the case of Minkowski space, the condition that the solutions be finite on the whole manifold requires the vanishing of the  $B_{lm}$ coefficients. This is because the Bessel function of type two $Y_l$ is divergent at
the origin. The flat space solutions are thus:
\begin{equation} \label{eq:bessel_j}
 f_{lm}(z)\, \sim \,j_{l}(z)
\end{equation}
\par
We observe that for our wormhole spacetime, the inferior limit of the radial coordinates is $r=a$. Thus, there is no point at which the functions $Y_{l}(z)$ diverge, which means they represent valid physical solutions.    
This being the case, it is recommended that we consider instead of the spherical Bessel function the corresponding Hankel function, which have the form: 
\begin{equation}
h^{(1,2)}_l(z) = \frac{e^{\mp iz}}{z} \left( ...  \right),
\end{equation}
where the paranthesis contains a polynomial with complex coefficients of degree $l$ in $1/z$. These functions describe ingoing and outgoing spherical waves. We build the solutions of the scalar field on the wormhole spacetime in terms of these
spherical Hankel functions. The radial solutions will have the following structure: waves incoming from infinity which scatter off the wormhole and result in a part reflected back towards infinity and a part transmitted through the wormhole.
There are two sets of modes:  
 \\ \par $1)$ waves ingoing from the region $\rho > 0$, indicated by the label $\sigma = +$ \par 
         $2)$ waves ingoing from the region $\rho < 0$, indicated by the label $\sigma = -$. \\ \\
The radial functions take the form:
\bq \label{eq:radial_function}
f^+_{\omega \ell}(\rho)=
\left\{ \begin{array}{rl}
N_{\omega \ell}\left[h^{(2)}_\ell(\omega r)+R_{\omega \ell}\, h^{(1)}_\ell(\omega r)\right]
\ \ \mbox{if}\, \rho>0\,\,
\\
N_{\omega \ell}\left[T_{\omega \ell}\, h^{(1)}_\ell(\omega r)\right]
\qquad\ \ \ \ \ \quad\quad \mbox{if}\, \rho<0,
\end{array}
\right. \\ \nn \\ \label{eq:radial_function2}
f^-_{\omega \ell}(\rho)=
\left\{ \begin{array}{rl}
N_{\omega \ell}\left[T_{\omega \ell}\, h^{(1)}_\ell(\omega r)\right]
\qquad\ \ \ \ \ \quad\quad \mbox{if}\, \rho>0\,\,
\\
N_{\omega \ell}\left[h^{(2)}_\ell(\omega r)+R_{\omega \ell}\, h^{(1)}_\ell(\omega r)\right]
\ \ \, \mbox{if}\, \rho<0.
\end{array}
\right.
\eq
The reflection and transmission coefficients can be found by imposing $\mathcal{C}^1$ class continuity conditions on the solutions at $\rho = 0$. The conditions can be written as:
\bq
\varphi(a+0^+)\, &=&\ \ \, \varphi(a+0^-) \nn \\
\varphi(a+0^+)'&=& -\varphi(a+0^-)',
\eq
where the derivatives are taken with respect to the coordinate $r$, and the minus sign in the second term arises because:
\begin{equation}
dr\, =\, sgn(\rho)\,d\rho
\end{equation}
After a bit of algebra and making use of the well known Wronskian formula for the spherical Hankel functions \cite{a27}: 
\begin{equation}
\mathcal{W}\,[h_l^{(1)}(z),h_l^{(2)}(z)] = - \frac{2i}{z^2},
\end{equation}
we arrive at the following expression for the reflection and transmission coefficients:
\begin{eqnarray} \label{eq:reflection_coef}
R_{\omega \ell} &=&\frac{[h^{(1)}_\ell(\omega a)\,
h^{(2)}_\ell(\omega a)]'}{2h^{(1)}_\ell(\omega a)\, h^{(1)}_\ell(\omega a)^\prime},
\\ \label{eq:transmission_coef}
T_{\omega \ell }&=&\frac{-i(a \omega)^2}{h^{(1)}_\ell(\omega a)\,
h^{(1)}_\ell(\omega a)^\prime}.
\end{eqnarray}
The explicit form of the coefficients for the first few values $l = 0,1,2$, is found in Appendix~\ref{AppendixA}.  
The normalization constant $N_{\omega l}$ can be found by imposing ortonormality on the modes (with respect to the usual scalar product):
\bq
\qquad (\varphi_{\omega\, \ell\, m}^\sigma,
\, \varphi^{\sigma^\prime}_{\omega' \ell{\,'} m'})
=
\delta_{\sigma,\, \sigma'}\, \delta(\omega-\omega')\,\delta_{\ell,\, \ell^{\,\prime}}\,\delta_{m,\, m'},
\eq
The resulting normalization constant is:
\begin{equation} \label{eq:normalization_coef}
N_{\omega \ell}=\frac{\omega}{\sqrt{2\pi}}
\end{equation}
\par

\begin{figure}
\centering
\begin{subfigure}{.5\textwidth}
  \centering
  \includegraphics[width=.95\linewidth]{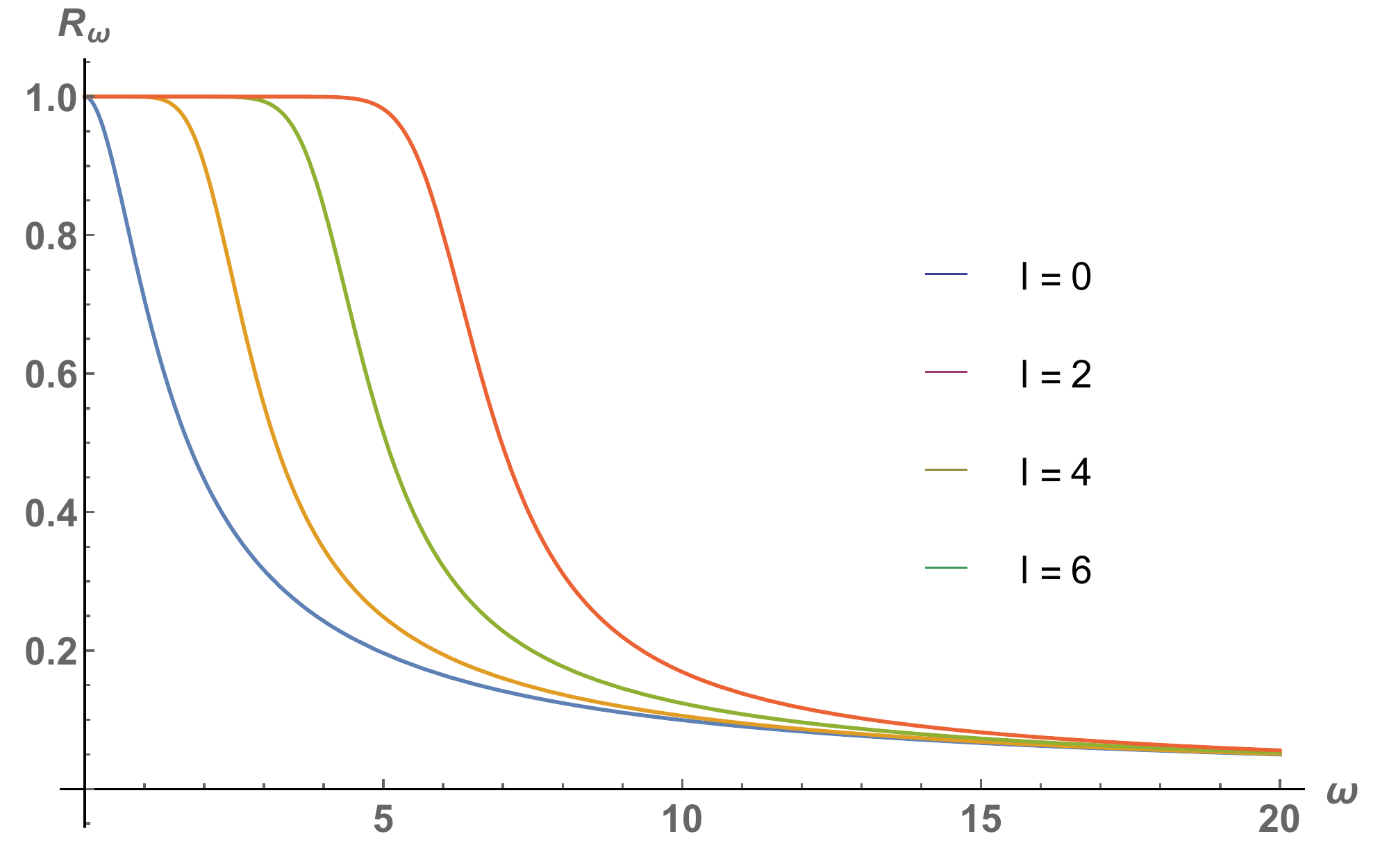}
  \label{fig:4.1a}
\end{subfigure}%
\begin{subfigure}{.5\textwidth}
  \centering
  \includegraphics[width=.95\linewidth]{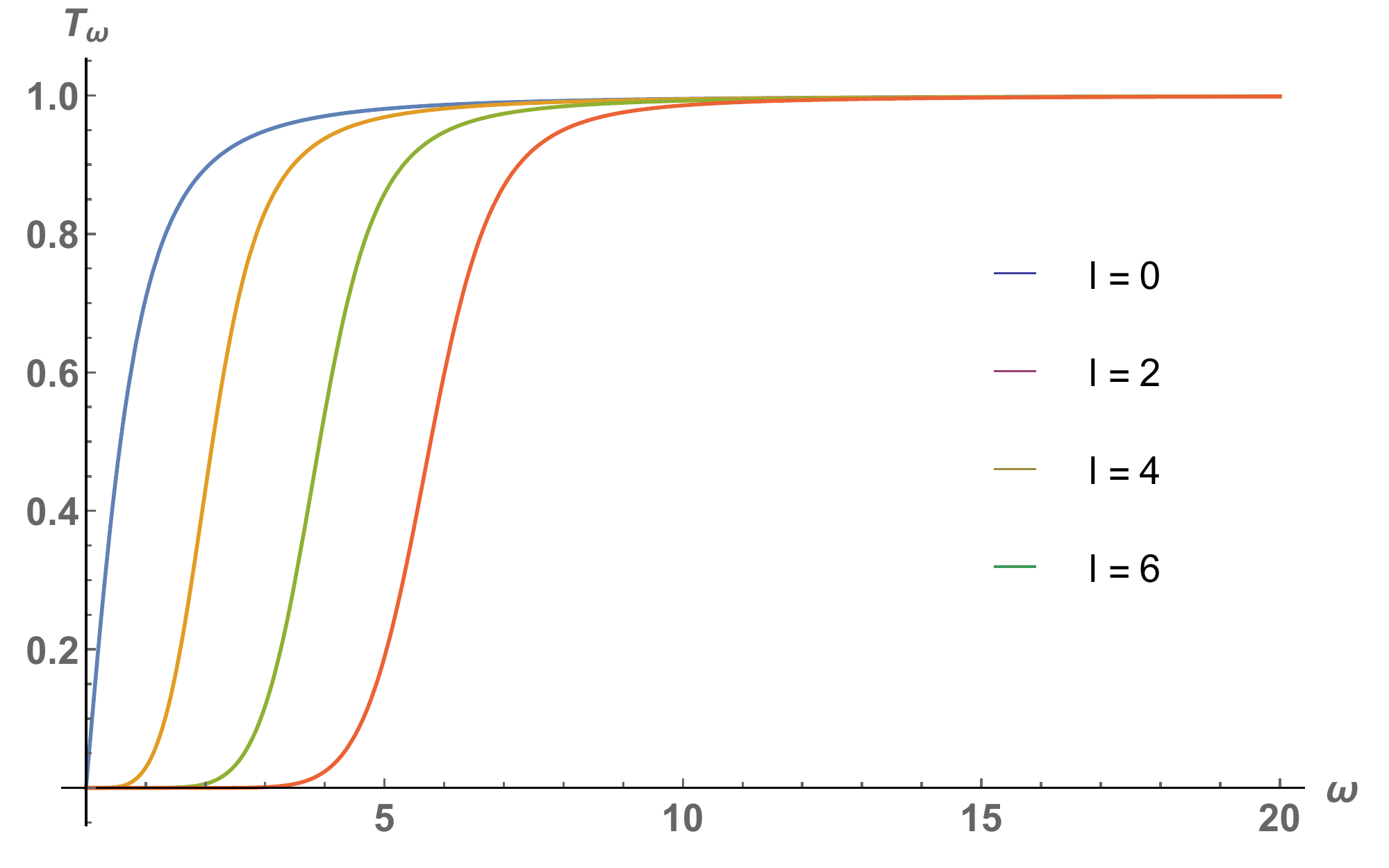}
   \label{fig:4.1b}
\end{subfigure}
\caption{\emph{The reflection $R_\omega$ and transmission $T_\omega$ coefficients for a wormhole with throat of unit radius $a = 1$.}}
\label{fig:4.2}
\end{figure}
\begin{figure}
\centering
\begin{subfigure}{.5\textwidth}
  \centering
  \includegraphics[width=.95\linewidth]{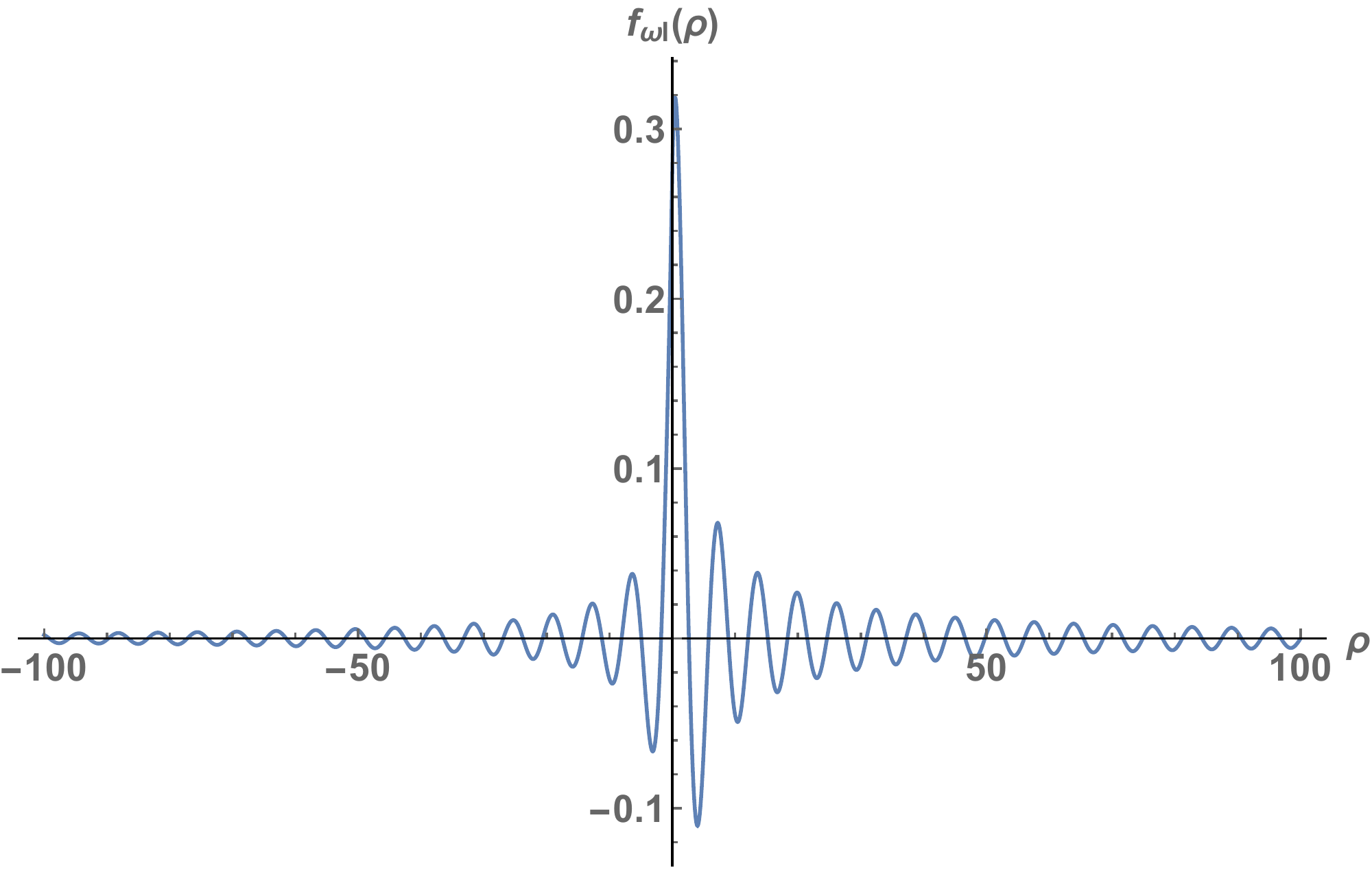}
  \label{fig:4.2a}
\end{subfigure}%
\begin{subfigure}{.5\textwidth}
  \centering
  \includegraphics[width=.95\linewidth]{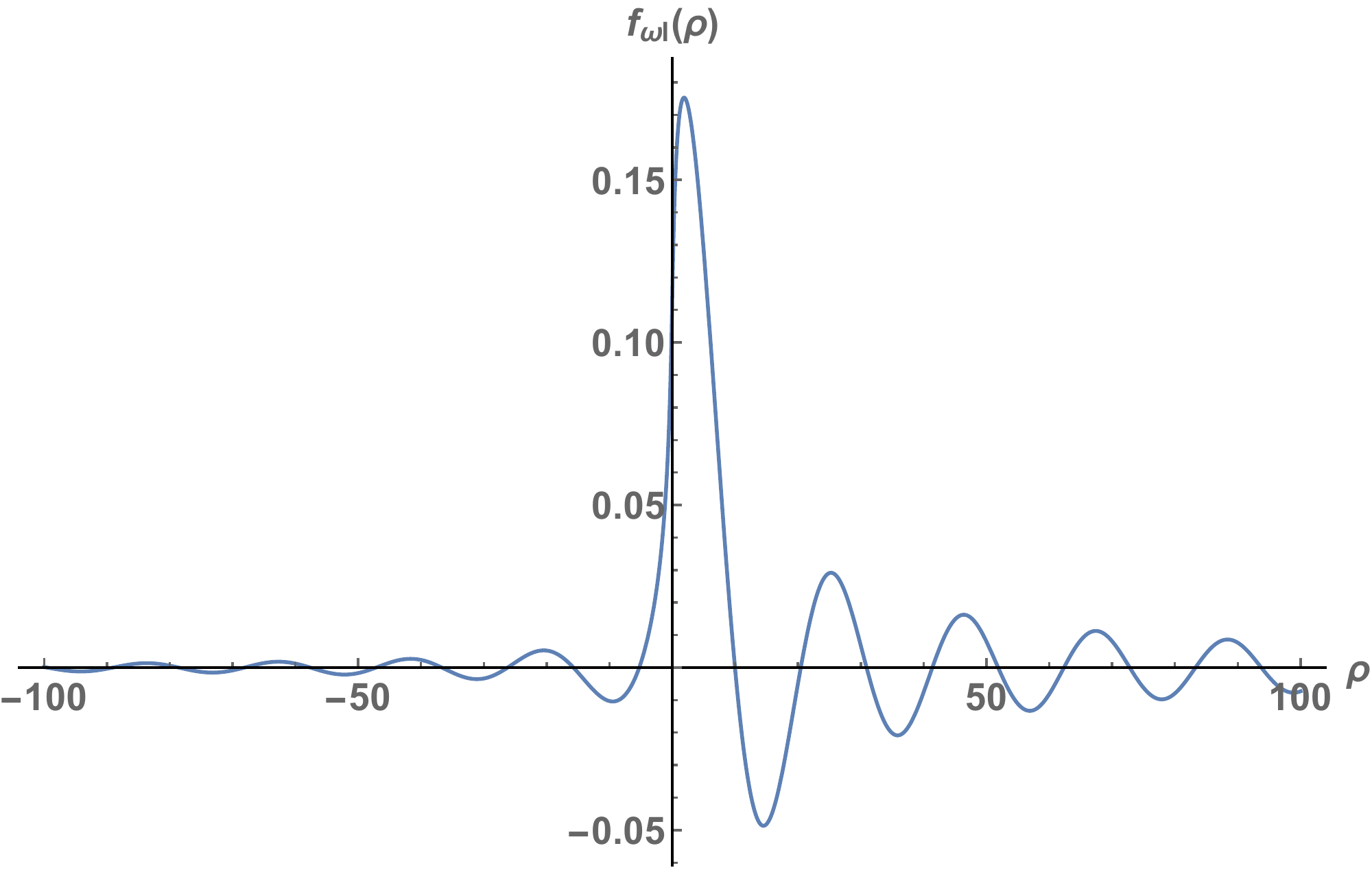}
   \label{fig:4.2b}
\end{subfigure}
\caption{\emph{The real part of the radial functions \eqref{eq:radial_function}, which are incident from the region $\rho > 0$.}}
\label{fig:4.3}
\end{figure}
The radial functions $f^{\pm}_{\omega l} (r)$ represent the wave incoming from the universe $\sigma = \pm$, that scatter off the wormhole, resulting in a reflected an a transmitted part. The modes that have wavelengths $1/\omega \ll a$
(where $a$ is the radius of the wormhole throat) ca not penetrate the wormhole and are mostly reflected. Obversely, the modes with $1/\omega \gg a$ largely pass through the wormhole unaffected, the scattered part being very small. This 
can be seen from the sudden transition in the reflection and transmission coefficients in Fig.~\ref{fig:4.2} and from the change in the complete radial wavefunctions as they pass the wormhole as shown in Fig.~\ref{fig:4.3}.
For small frequencies the reflection coefficients have unit value while the transmission coefficient vanishes (total reflection). As we increase the frequency, there is a very fast shift in behavior of the coefficients, the reflection 
coefficient falling to zero, while the transmission coefficient commutes to unity (complete transmission). \par
In the following, we apply the canonical quantification proceduce for the scalar field $\varphi$. The mode expansion of the field is:

\bq \label{eq:mode_expansion_re}
\varphi ({\bf x})=
\sum\limits_{i}^{}\{a_i \varphi_i({\bf x})+a^\dag_i \varphi^*_i ({\bf x})\},
\eq
where the index $i$ denotes the complet set of quantum number necessary to uniquely describe the modes, which in our case are $i \equiv \{\sigma,\omega,l,m\}$. The coefficients $a^\dag$ and $a$ become creation and annihilation
operators, that obey the commutation relations:
\begin{equation}
[a_{\alpha},a^\dag_{\beta}] = \delta_{\alpha,\beta}\ ,\ \ \ [a_{\alpha},a_{\beta}] = [a^\dag_{\alpha},a^\dag_{\beta}] = 0,
\end{equation}
where $\delta_{\alpha,\beta}$ is the Kroenecker delta function in the case of discrete indices, and the Dirac delta function for continuous indices. \par
The quantunm modes $\varphi_\alpha(x)$ are solutions to the Klein-Gordon equation \eqref{eq:helmholtz}, and have the form:
\bq \label{eq:modes_explicit}
\varphi^\sigma_{\omega \ell\, m}(t,\rho,\theta,\phi) =
f^\sigma_{\omega \ell}\,(r)\, Y_{\ell\, m}(\theta,\phi)\ e^{-i\omega t},
\eq
with $Y_{\ell\, m}$ being the spherical harmonics defined by eq.~\eqref{eq:spherical_harmonics}, the radial functions $f_{\omega \ell}$ are given by \eqref{eq:radial_function} and \eqref{eq:radial_function2}, 
along with the reflection and transmission coefficients \eqref{eq:reflection_coef} and \eqref{eq:transmission_coef}, and with the normalization  \eqref{eq:normalization_coef}. 

\section{Transition probability and rate}
The quantities that we wish to find are the transition probability and rate of the detector, from the ground state $\vert E_0\rangle$ to the excited state $\vert E\rangle$, as it evolves on the trajectory. The first step is to 
write down the transition amplitude of the system formed by the detector and the scalar field from the state $\vert 0, E_0\rangle =\vert 0\rangle \vert E_0\rangle$, i.e. the ground state of the detector and the vacuum state of the field,
to the state $\vert \alpha, E\rangle = \vert\varphi_\alpha \rangle\vert E\rangle$, i.e. the excited state of the detector and the state of the field with the quantum numbers $\alpha$. The amplitude is given by:
\bq 
{\cal A}_{\,0,\, E_0\rightarrow \,\alpha,\, E}(\tau)=
\langle E \vert \mu(0)\vert E_0\rangle
\times
\int\limits_{\tau_0}^{\tau} d\tau\ e^{iE\tau}
\langle \alpha|\varphi(x(\tau))|0\rangle
\equiv  A_\alpha(E,\tau).
\eq
We neglect the first term in the amplitude, the detector ''sensitivity``. As we have noted earlier, this quantity depends only on the internal detail of the detector and not on the the trajectory. In order to separate the contribution from the 
trajectory, we fix the sensitivity to unity. Using the mode expansion \eqref{eq:mode_expansion_re} of the field, the amplitude becomes:
\bq \label{eq:amplitude}
A_\alpha(E,\tau)=
\int\limits_{\tau_0}^{\tau} d\tau\  e^{iE\tau}\varphi^*_\alpha(x(\tau)),
\eq
We observe that the amplitude is non-zero only in the case that $\alpha$ describes a one-particle state. Given the amplitude \eqref{eq:amplitude}, the transition probability is obtained as:
\begin{equation} \label{eq:probability_partial}
\mathcal{P}_\alpha (E,\tau) = \vert A_\alpha(E,\tau) \vert ^{\,2}
\end{equation}
The physical quantity of interest is the probability (and rate) that the detector suffers a transition from the ground state to the excited state, indifferent to what the final state of the scalar field is. In order to obtain this ''total``
probability, we have to sum (average to be more precise) the ''partial`` probabilities \eqref{eq:probability_partial} over all final states of the field, i.e. sum and integrate over all quantum numbers $\alpha$. The total probability is then
written as:
\begin{equation}
\mathcal{P}(E,\tau) = \sum_\alpha \mathcal{P}_\alpha (E,\tau)
\end{equation}
The corresponding transition rate can be obtained directly from the final probability as:
\begin{equation} \label{eq:rate_v1}
\mathcal{R}(E,\tau) = \frac{d}{d\tau} \mathcal{P} (E,\tau)
\end{equation}
Equation \eqref{eq:rate_v1} should be interpreted in the manner described in the paragraph following eq.~\eqref{eq:physical_rate}. \\
This above procedure is useful when the Wightman functions is known:
\begin{equation}
W(x,x')= \sum\limits_{\alpha}^{} \varphi_\alpha(x)\varphi^*_\alpha(x'),
\end{equation}
and with its help we can write the total probability as:
\bq
\mathcal{P}(E,\tau)=
\int\limits_{-\infty}^{\tau}d\tau_1 \int\limits_{-\infty}^{\tau} d\tau_2\,\,W(x(\tau_1),x(\tau_2)).
\eq
An alternative approach, in the case that we don't know the explicit form of the correlation function, is to obtain the transition rate by first evaluating the ''partial rates`` derived from the partial probabilities 
\eqref{eq:probability_partial}:
\bq
\mathcal{R}_\alpha(E,\tau)=\frac{d}{d\tau}\, \mathcal{P}_\alpha(E,\tau)
\eq
\bq
\mathcal{R}_\alpha(E,\tau)=2\mbox{Re}
\left[ \varphi_\alpha^*(x(\tau)) \int\limits_{-\infty}^{\tau}
d\tau'  e^{-iE(\tau-\tau^\prime)}
\varphi_\alpha(x(\tau'))\right],
\eq
and then performing the summation over the final states of the field:
\begin{equation}
\mathcal{R}(E,\tau)= \sum_{\alpha} \mathcal{R}_\alpha(E,\tau)
\end{equation}
In our case the set of quantum numbers that uniquely describe the modes are $\alpha = \{\sigma,\omega,l,m\}$, and thus:
\begin{equation}
\sum_\alpha\ \ \ \rightarrow\ \ \ \int\limits_{0}^{\infty}d\omega\ \sum_l\ \sum_m\ \sum_\sigma,
\end{equation}
while the modes $\varphi_{\alpha}$ are given by eq.\eqref{eq:modes_explicit}.

\section{Inertial trajectories in Minkowski space}
First, we look at the response of an inertial detector in Minkowski space as presented in the previous chapter, but now in spherical coordinates and using the spherical modes \eqref{eq:bessel_j}. The full Minkowskian solutions
in terms of spherical modes are:
\begin{equation}
 \varphi_{\omega l m}(r,\theta,\phi) = \sqrt{\frac{\omega}{\pi}} j_l(\omega r)Y_{lm}(\theta,\phi)e^{-i\omega t}
\end{equation}
Without lack of generality we can consider only trajectories in the plane $\theta = \pi/2$. This selects the $m=0$ mode, and the spherical harmonics reduce to:
\begin{equation}
 Y_{l0} = \sqrt{\frac{2l+1}{4\pi}}
\end{equation}
With this observation, the amplitude for a detector on an arbitrary trajectory $t = t(\tau), r = r(\tau)$ is written as:
\begin{eqnarray}
 \mathcal{A}_{\omega l}^M(E,\tau) &=& \sqrt{\frac{\omega}{\pi}}\int_{-\infty}^{\tau}d\tau e^{iE\tau}\, j_l(\omega r) Y_{lm}(\theta \phi) e^{i\omega t} \\
 &=& \sqrt{\frac{(2l+1)\omega}{4\pi^2}}\int_{-\infty}^{\tau}d\tau e^{i(E\tau+\omega t(\tau))} j_l(\omega r(\tau))   
\end{eqnarray}
\subsubsection{\emph{Static detector}} 
As we have seen, the probability for an eternally static detector on the trajectory $\{ r = const, t = \tau \}$ vanishes. If we do the calculation with spherical modes, the partial amplitudes are:
\begin{eqnarray}
 \mathcal{A}_{\omega l}^M(E,\tau) &=& \sqrt{\frac{(2l+1)\omega}{4\pi^2}}\, j_l(\omega r)\int_{-\infty}^{\infty}d\tau e^{i(E+\omega)\tau} \\
 &=& \sqrt{(2l+1)\omega} \,j_l(\omega r) \delta(E + \omega)
\end{eqnarray}
The quantity in the bracket is always positive, and thus the Dirac delta function enforces the vanishing of the transition amplitude at the level of each spherical mode. 

\subsubsection{\emph{Inertial motion}}
Next we turn to the case of an inertial detector, evolving on the trajectory:
\begin{eqnarray} \label{eq:inertial_spherical_trajectory}
 t &=& \gamma \tau_0 \\
 r &=& \vert v\gamma \tau \vert \\
 \theta &=&  \Theta(\tau)\frac{\pi}{2} - \Theta(-\tau) \frac{\pi}{2} \\
 \phi &=& \Theta(\tau) \pi  
\end{eqnarray}
The response of the detector is easiest to obtain by looking directly at the probability:
\begin{eqnarray} \label{eq:inertial_mink_spherical}
 P_{\omega} &=&  \sum_{l=0}^{\infty} \frac{(2l+1)\omega}{4\pi^2} \left\vert \int\limits_{-\infty}^{\infty} d\tau e^{i(E+\gamma \omega)\tau} j_l(\omega r) \right\vert ^2 \\
 &=& \frac{\omega}{4\pi^2} \int\limits_{-\infty}^{\infty} \int\limits_{-\infty}^{\infty} e^{i(E+\gamma \omega)(\tau-\tau')} \sum_{l=0}^{\infty}(2l+1) j_l(\omega r) j_l(\omega r) \\
 &=& \frac{\omega}{4\pi^2} \int\limits_{-\infty}^{\infty} \int\limits_{-\infty}^{\infty} e^{i(E+\gamma \omega)(\tau-\tau')} \\
 &=& \omega\, \delta(E+\gamma \omega)^2
\end{eqnarray}
In the intermediate steps we have exploited the parity of the spherical Bessel functions $j(-z) = (-1)^l j(z)$ and used the relation \cite{a27}:
\begin{equation} 
 \sum_{l = 0}^{\infty} (2l+1) j_l(z) j_l(z) = 1.
\end{equation}
Another tacit assumption we have made is that all sums and integrals converge, so that we can commute the integration and summation operations in \eqref{eq:inertial_mink_spherical}. The fact that the results for 
a static and arbitrary inertial trajectory coincide is a sign that we are on track. Indeed this is guaranteed by the fact that the Wightman function is boost invariant. 

\subsubsection{\emph{Finite-time trajectory}}
We turn now to the case when the detector is coupled on at $\tau_0$, but the measurement is taken at a finite time. The trajectory is inertial and is described by \eqref{eq:inertial_spherical_trajectory}, but
now the integration is done between the finite times $\tau_0$ and $\tau$. For simplicity we look at the contribution of the $l=0$ mode to the response of the detector. In this case the amplitude is written as:
\begin{equation}
 A^{M}_{\omega 0}(E,\tau) = \frac{\sqrt{\omega}}{2\pi} \int_{\tau_0}^\tau d\tau\, e^{i(E+\gamma \omega)\tau} j_0(\omega r).
\end{equation}
Taking account of the parity of the spherical Bessel functions, we can expand the integrand as follows:
\begin{eqnarray}
 e^{i(E+\gamma \omega)\tau} j_0(\omega r) &=& e^{i(E+\gamma \omega)\tau}\left(\frac{e^{i\omega v \gamma \tau} - e^{-i\omega v \gamma \tau}}{2i\omega v \gamma \tau}\right) \\
 &=& \frac{1}{2i\omega v \gamma}\left(\frac{e^{i\alpha_+\tau} - e^{i\alpha_-\tau}}{\tau} \right),
\end{eqnarray}
where we have introduced the notation:
\begin{equation} \label{eq:alpha}
 \alpha_{\pm} = E + \gamma\omega(1\pm v) 
\end{equation}
The amplitude then becomes:
\begin{eqnarray} \label{eq:amp_mink_spherical_w0}
 A^{M}_{\omega 0}(E,\tau) &=& \frac{1}{4\pi i} \frac{1}{\sqrt{\omega} v \gamma} \int_{\tau_0}^\tau \frac{d\tau}{\tau} \left(e^{i\alpha_+\tau} - e^{i\alpha_-\tau}\right) \\
 &=&  \frac{1}{4\pi i} \frac{1}{\sqrt{\omega} v \gamma} \left[\text{Ei}(i\alpha_\text{\tiny +}\tau) - \text{Ei}(i\alpha_\text{\tiny +}\tau_0) + \text{Ei}(-i\alpha_\text{\small -}\tau) - \text{Ei}(-i\alpha_\text{\small -}\tau_0)\right]. \nn
\end{eqnarray}
The exponential integral functions $\text{Ei}$ is defined as in \eqref{eq:exponential_integral}. \par
Notice that these amplitudes are in general time-dependent quantities. On the other hand, it can be checked that if we set $\tau_0 = -\tau_0$ they vanish identically, independent of the value of $\omega$. 
In particular, the result is zero for $\tau = -\tau_0 \rightarrow \infty$. If we keep $\tau$ finite, and take the limit $\tau_0 \rightarrow -\infty$ we have
\begin{equation}
\lim_{\tau_0 \rightarrow -\infty} \text{Ei}(\pm i\tau_0) = \pm i\pi.
\end{equation}
Thus, the amplitude \eqref{eq:amp_mink_spherical_w0} becomes:
\begin{equation} \
  A^{M}_{\omega 0}(E,\tau) = \frac{1}{4\pi i} \frac{1}{\sqrt{\omega} v \gamma} \left[\text{Ei}(i\alpha_\text{\tiny +}\tau) + \text{Ei}(-i\alpha_\text{\small -}\tau) \right]. 
\end{equation}

\begin{figure}
\centering
\begin{subfigure}{.5\textwidth}
  \centering
  \includegraphics[width=.95\linewidth]{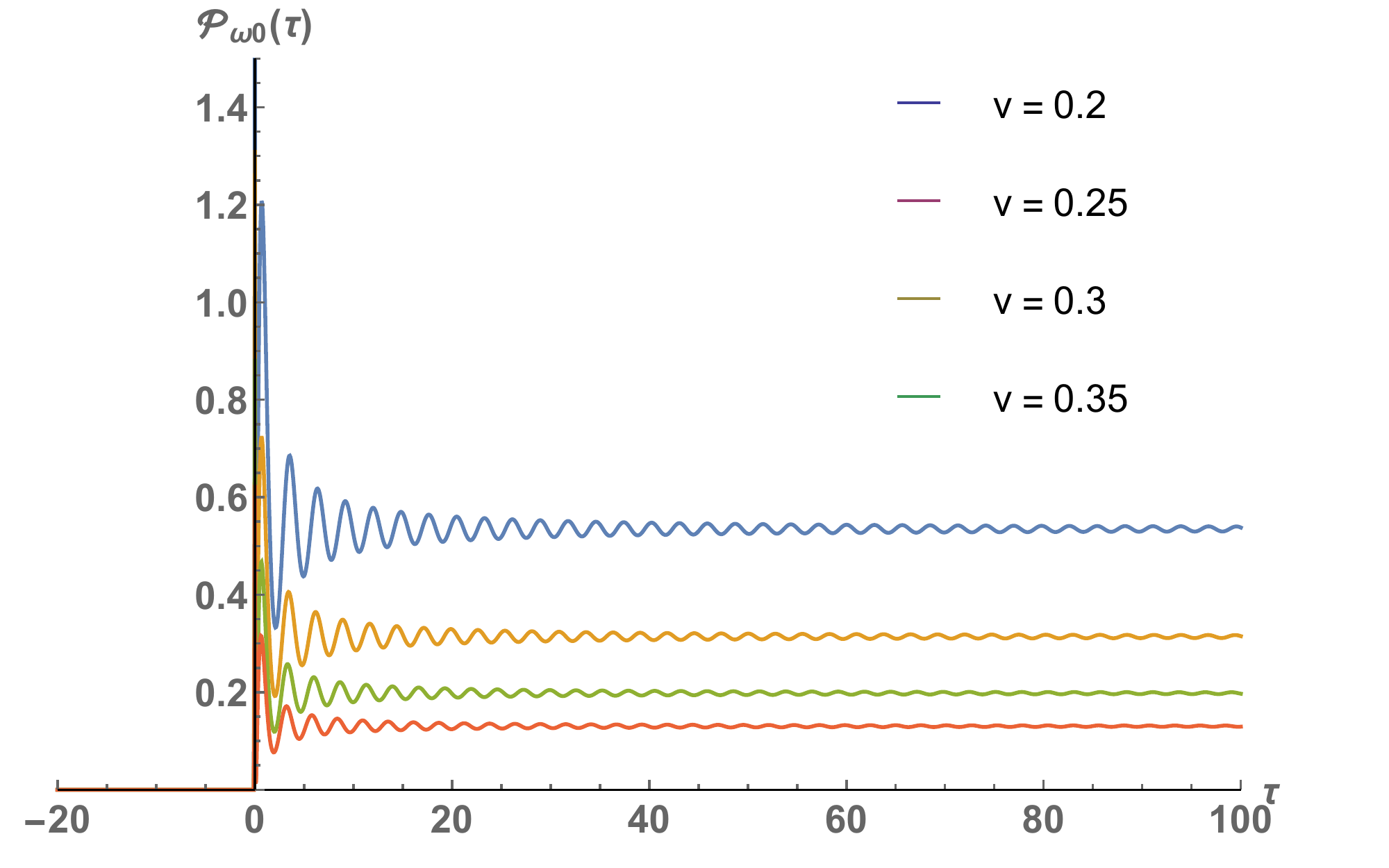}
  \label{fig:4.2a}
\end{subfigure}%
\begin{subfigure}{.5\textwidth}
  \centering
  \includegraphics[width=.95\linewidth]{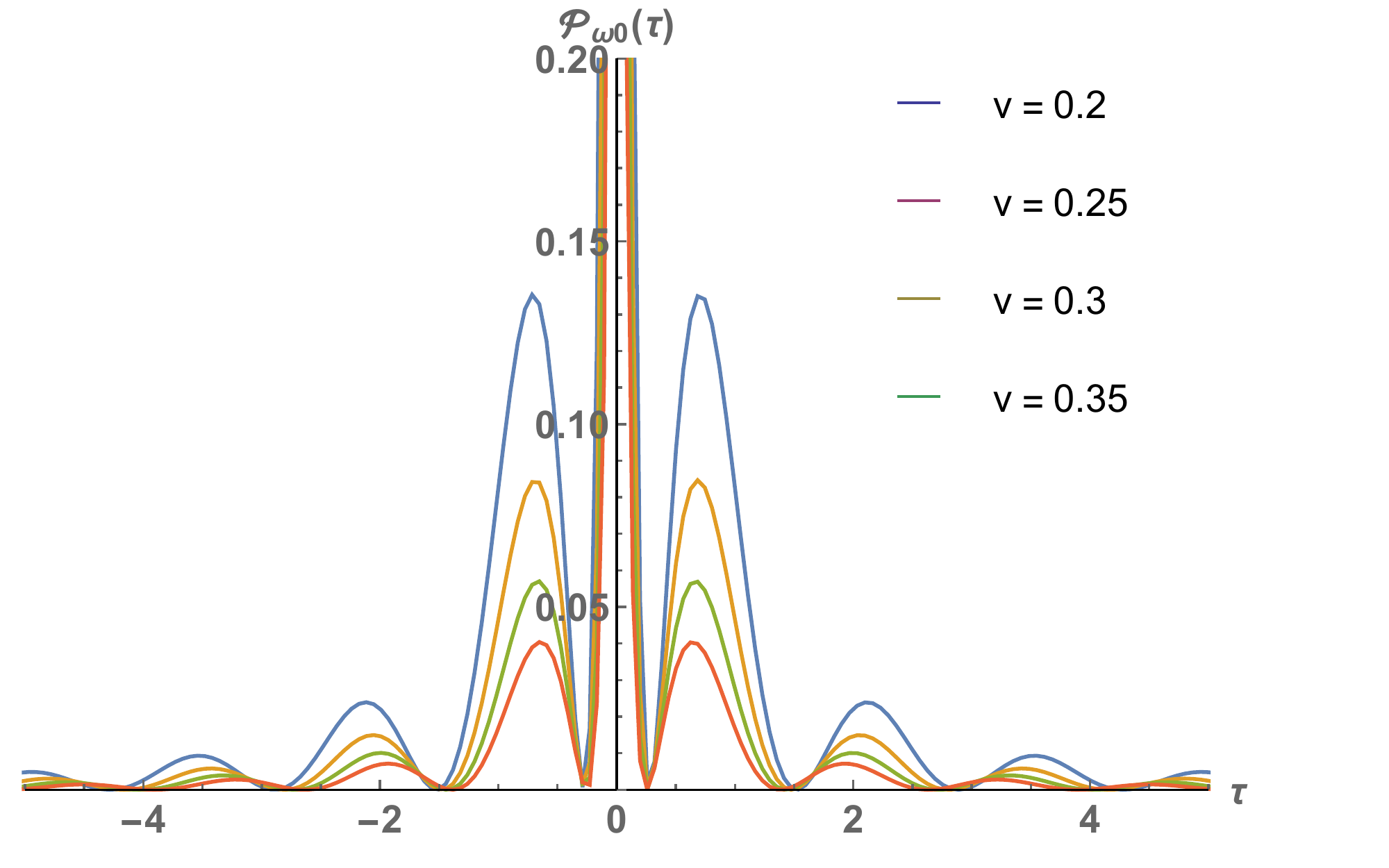}
   \label{fig:4.2b}
\end{subfigure}
\caption{\emph{S-wave contribution to the transition probability of an inertial detector. When the detector is switched on at a finite time $\tau_0 = -0.1$ (left), the probability at large times stabilizes at a finite value, 
while in the case of coupling at $\tau = -\infty$ they vanish. In both cases the partial probability shows a complicated time dependence. The curves represent different velocities, and have the parameters $E=1, \omega = 1$. }}
\label{fig:4.4}
\end{figure}
Fig.~\ref{fig:4.4} shows the s-wave partial probability (modulus squared of \eqref{eq:amp_mink_spherical_w0}) for the case of finite and infinite-time coupling. In both case the probabilities are rapidly varying functions of time. 
In the infinite time limit in both cases the probability tend to a constant value. While for an eternally copled detector the probability vanishes at large times, for the case of a finite-time coupling it tends to a positive nonzero value. \par
To obtain the s-wave contribution to the total probability, we need to integrate over all frequencies:  
\begin{equation} \label{eq:prob_mink_l0}
 \mathcal{P}_{0} = \frac{1}{16\pi^2v^2\gamma^2} \int_{0}^{\infty} \frac{d\omega}{\omega} \ \left[\text{Ei}(i\alpha_\text{\tiny +}\tau) - \text{Ei}(i\alpha_\text{\tiny +}\tau_0) + \text{Ei}(-i\alpha_\text{\small -}\tau) - \text{Ei}(-i\alpha_\text{\small -}\tau_0)\right]^2
\end{equation}
The result of the integral \eqref{eq:prob_mink_l0} turns out to be a divergent quantity in general. Only in the special case $\tau = -\tau_0$ does attain the result expected on physical grounds, which is that the detector does not get 
excited. If the measurement is taken at suffieciently large times after the detector has been switched on then the probability goes to a constant value, and the rate vanishes. However, only in the limit $\tau = -\tau_0 \rightarrow \infty
\ \text{will}\  P_{\omega 0} \rightarrow 0$.  \par
These probabilities with the arising divergences, have been the subject of a thorough investigation in Ref.\cite{a28}.

\section{Inertial trajectories in a WH spacetime}
Now we turn the case of the wormhole spacetime. The full solution for the Klein-Gordon equation in the wormhole metric \eqref{eq:wh_metric} is:
\begin{equation}
 \varphi^\sigma_{\omega l m}(\omega,r,\theta,\phi) = \sqrt{\frac{\omega}{2\pi}} f^\sigma_{\omega l}(r) Y_{lm}(\theta,\phi)e^{-i\omega t}
\end{equation}
As before, we consider only trajectories that are in the plane $\theta = \pi/2$, which reduces the spherical harmonics to the simple form:
\begin{equation}
 Y_{lm} = \sqrt{\frac{2l+1}{4\pi}}.
\end{equation}
The transition amplitude for a general trajectory $r = r(\tau), t = t(\tau)$, is written as:
\begin{equation} \label{eq:amp_wh_general}
A^\sigma_{\omega,l}(E,a,\tau) = \sqrt{\frac{\omega}{8\pi^2}}\int_{\tau_0}^\tau d\tau  \left[f^\sigma_{\omega l}(r)\right]^* e^{i(E\tau+\omega t)}
\end{equation}
The substantial difference in this case is that we have two sets of mode functions \eqref{eq:radial_function} and \eqref{eq:radial_function2}, distinguished by the label $\sigma$. Lorentz invariance is in this case obviously broken.
Notice that the trajectory which is eternally static can in no way be tranformed via a boost into an inertial trajectory which crosses the wormhole into the second universe. Intuition says then that the static detector should
still have vanishing probability, while the inertial trajectory through the wormhole should (or at least could) have a nonvanishing response. Lets look again at these trajectories separately.
\subsubsection{\emph{Static detector}}
Consider again a detector which is eternally static and is coupled at all times ($\tau_0 \rightarrow -\infty$). Inserting the trajectory $r = const., t = \tau$ into the amplitude \eqref{eq:amp_wh_general} results in:  
\begin{eqnarray}
 A^\sigma_{\omega,l}(E,a,\tau) &=& \sqrt{\frac{\omega}{8\pi^2}} \left[f^\sigma_{\omega l}(r)\right]^* \int_{-\infty}^\infty d\tau e^{i(E+\omega)\tau} \\ 
 &=& \sqrt{\frac{\omega}{2}}  \left[f^\sigma_{\omega l}(r)\right]^* \delta(E+\omega),
\end{eqnarray}
which is zero as expected. 
\subsubsection{\emph{Inertial motion}}
Let us now turn to the most interesting case, that of an inertial trajectory which crosses the wormhole. The trajectory is given by:
\begin{eqnarray} \label{eq:trajectory_wh}
t(\tau) &=& \gamma \tau \nonumber \\
r(\tau) &=& a +\vert \gamma   v\tau \vert \nonumber \\
\phi(\tau) &=& 0 \\
\theta(\tau) &=& 0 \nonumber
\end{eqnarray}
The trajectory describes a uniform radial motion, with the detector incoming from $\rho \rightarrow \infty$ (the universe denoted with $\sigma = +$), which in the moment $\tau = 0$ passes through the throat of the wormhole
and then continues to move uniformly radially outwards to infinity through the other universe (denoted with with $\sigma= -$), as depicted by the arrows in Fig.~\ref{fig:4.1}. Due to the non-trivial topology introduced by 
the existence of the wormhole, at the passage through $\rho = 0$ the angular coordinates transform as:
\begin{eqnarray}
\phi \rightarrow \phi   \nonumber, \qquad 
\theta \rightarrow \theta
\end{eqnarray}
Note that this trajectory is continuous. In the case of uniform linear motion in Minkowski space, at the passage through the origin at $\rho = 0$, the angular coordinates transform as:
\begin{eqnarray}
\phi \rightarrow \phi + \pi \nn \qquad 
\theta \rightarrow \pi - \theta .
\end{eqnarray}
The trajectory \eqref{eq:trajectory_wh} describes a detector on uniform linear motion, with velocity $v$, up untill it reaches the throat at the moment $\tau = 0$, point at which it passes through the wormhole and continues 
back on the same trajectory in the second universe, with velocity $-v$, as indicated by the arrows in Fig.~\ref{fig:4.1}.\\
The transition amplitude for the case $\tau < 0$ has the following form:
\begin{eqnarray} \label{eq:amp_wh_inertial}
A^+_{\omega l}(E,\tau) & = & \int\limits_{-\infty}^{\tau} d\tau\, e^{iE\tau} \left(\,f^+_{\omega l} (r)\, Y_{l0}(\phi,\theta) \,e^{-i\omega t	}\right)^* \nn\\
&=& \sqrt{\frac{(2l+1)\omega}{8\pi^2}}\int\limits_{-\infty}^{\tau} d\tau \left( h_l^{(1)}(\omega r) - R^{\,*}_{\omega l}\, h_l^{(2)}(\omega r)\right)\,e^{i(E\,+\,\gamma \omega)\tau} \nn \\
A^-_{\omega l}(E,\tau) & = & \int\limits_{-\infty}^{\tau} d\tau\, e^{iE\tau} \left(\,f^-_{\omega l} (r)\, Y_{l0}(\phi,\theta) \,e^{-i\omega t	}\right)^* \nn\\
&=& \sqrt{\frac{(2l+1)\omega}{8\pi^2}}\int\limits_{-\infty}^{\tau} d\tau \left( T^{\,*}_{\omega l}\, h_l^{(2)}(\omega r)\right)\,e^{i(E\,+\,\gamma \omega)\tau} \nn
\end{eqnarray}
As in the flat space case, it is best if we look directly at the transition probability:
\begin{eqnarray} \label{eq:prob_wh_Mink+WH}
 \mathcal{P}_{\omega l} &=& \left\vert A^+_{\omega l}\right\vert^2 + \left\vert A^-_{\omega l}\right\vert^2 \\
   &=& \frac{(2l+1)\omega}{8\pi^2}\int\limits_{-\infty}^{\infty}d\tau \int\limits_{-\infty}^{\infty}d\tau' e^{i(E+\gamma \omega)(\tau-\tau')} \left\{  \left( h_l^{(1)}(\omega r) + R^{\,*}_{\omega l}\, h_l^{(2)}(\omega r)\right) \nn \right. \\
  &&  \times  \left. \left( h_l^{(2)}(\omega r') + R_{\omega l}\, h_l^{(1)}(\omega r')\right)  + \left\vert T_{\omega l}\right\vert^2  h_l^{(2)}(\omega r)h_l^{(1)}(\omega r') \right\} \nn \\
  &=& \frac{(2l+1)\omega}{8\pi^2}\int\limits_{-\infty}^{\infty}d\tau \int\limits_{-\infty}^{\infty}d\tau' e^{i(E+\gamma \omega)(\tau-\tau')} \nn \\
  && \times \left\{ 4j_l(\omega r)j_l(\omega r') + 2\Re \left[(R_{\omega l} - 1) h^{(1)}_l(\omega r)h^{(1)}_l(\omega r')\right]\right\} \nn \\
  &\equiv& \mathcal{P}^M_{\omega l} + \Delta \mathcal{P}_{\omega l} \nn
\end{eqnarray}
In the intermediate steps we have used the fact that the two Hankel functions are complex conjugates of each other :
\begin{equation}
 \left(h^{(1)}_{l}(z)\right)^* = h^{(2)}_{l}(z), \qquad \left(h^{(2)}_{l}(z)\right)^* = h^{(1)}_{l}(z),
\end{equation}
as well as the relation between the reflection and transmission coefficients:
\begin{equation}
 \vert T_{\omega l} \vert^2 + \vert R_{\omega l}\vert^2 = 1
\end{equation}
The probability can be separated into a Minkowskian contribution and a correction. The term denoted with $\mathcal{P}^M_{\omega l}$ can be seen to be truly equivalent to the flat space contribution if we change the integration 
variable to: 
\begin{equation}
 \eta = \tau + sgn(\tau)a/(v\gamma),
\end{equation}
such that
\begin{eqnarray}
r \equiv a + v\gamma \vert \tau \vert &=& sgn(\tau) v\gamma  \eta,
\end{eqnarray}
and $dt = d\eta$. \par
As we have ween previously, the flat space contribution vanishes for the case of an inertial trajectory. The probability in the case of the wormhole spacetime, the probability is completally determined by the  correction
term $\Delta \mathcal{P}_{\omega l}$. We write it out explicitly for the s-wave case: 
\begin{eqnarray}
 \Delta \mathcal{P}_{\omega 0} &=& \frac{\omega}{4\pi^2}\int\limits_{-\infty}^\infty d\tau \int\limits_{-\infty}^\infty d\tau'\, e^{i(E+\gamma\omega)(\tau-\tau')}\,\Re \left[(R_{\omega 0} - 1) h^{(1)}_0(\omega r)h^{(1)}_0(\omega r')\right] \\
 &=& \frac{\omega}{4\pi^2}\int\limits_{-\infty}^\infty \frac{d\tau}{r} \int\limits_{-\infty}^\infty \frac{d\tau'}{r'}\, e^{i(E+\gamma\omega)(\tau-\tau')}\, \Re \left[ \left(1 -  \frac{e^{-2i\omega a}}{1-i\omega a} \right)\frac{e^{i\omega(r+r')}}{\omega^2rr'} \right] \nn\\
 &=& \frac{1}{4\pi^2\omega} \int\limits_{-\infty}^\infty \frac{d\tau}{r} \int\limits_{-\infty}^\infty \frac{d\tau'}{r'} e^{i(E+\gamma\omega)(\tau-\tau')}  \nn\\ 
 && \times \left( \cos \omega(r+r')  - \frac{\cos \omega(r+r'-2a)-\omega a\sin\omega(r+r'-2a)}{1-(\omega a)^2}  \right) \nn
\end{eqnarray}
Some conclusions can be readily drawn from the above form of the transition probability. First of all it has to be a positive quantity. Second, the probability vanishes in the limit $\omega a \rightarrow 0$, the first order term being
\begin{eqnarray} \label{eq:prob_wh_small_a}
 \mathcal{P}^{(1)}_{\omega 0} &\simeq& -\frac{a}{4\pi^2}\int\limits_{-\infty}^\infty \frac{d\tau}{r} \int\limits_{-\infty}^\infty \frac{d\tau'}{r'}  e^{i(E+\gamma\omega)(\tau-\tau')} \sin\omega(r+r') \\
 &=& -\frac{a}{4\pi^2}\ \Im\left( \int\limits_{-\infty}^\infty \frac{d\tau}{r} e^{i(E+\gamma\omega)\tau+i\omega r)}\int\limits_{-\infty}^\infty \frac{d\tau'}{r'} e^{-i(E+\gamma\omega)\tau+i\omega r)}\right) \nn \\
 &=& -\frac{a}{4\pi^2 v^2\gamma^2}\ \Im \left\{\left[ e^{-i\mathcal{E} a} \Gamma\left(0,-i\alpha_\text{\tiny +} a/v\gamma\right) - e^{i\mathcal{E} a} \Gamma\left(0,i\alpha_\text{\small -} a/v\gamma\right)  \right]^2\right\},
\end{eqnarray}
where $\Im$ represents the imaginary part of the bracketed expression, $\alpha_\text{\tiny +}$ and $\alpha_\text{\small -}$ are defined as in \eqref{eq:alpha}, and we have introduced the notation 
$\mathcal{E} = \left(E + \gamma \omega\right)/(v\gamma)$. The incomplete gamma function $\Gamma$ is defined as:
\begin{equation}
 \Gamma(s,x) =  \int_{x}^\infty t^{s-1} e^{-t}
\end{equation} 
We can further expand \eqref{eq:prob_wh_small_a} for small values of the wormhole radius $a$. Collecting the smallest order term in the expansion, we obtain:   
\begin{eqnarray}
  \mathcal{P}_{\omega 0}(a\rightarrow 0) &=& \frac{a}{2\pi v^2\gamma^2}\ln\left(\frac{\alpha_\text{\tiny +}}{\alpha_\text{\small -}}\right).
\end{eqnarray} 
The symbol $\hat{\gamma}$ represents the Euler-Mascheroni constant.
\subsection{The contribution of the $l = 0$ mode}

We investigate the response of the Unruh-deWitt detector moving on the trajectory \eqref{eq:trajectory_wh} through the wormhole spacetime described by the metric \eqref{eq:wh_metric}. In the previous section we have calculated 
the response of the detector in the future infinite limit. Now we look at the transition probability as a function of time. We expect that the detector will present a non-trivial behavior as it passes through the wormhole,
features which should dissapear as we go towards large future times, with the probability stabilizing to a constant, but non-zero, value. \par 
We start the analysis by looking at the contribution of the $l = 0$ term in the angular expansion, oftern called the s-wave, to the total transition probability and rate. The analysis of the s-wave mode is significant for 
several reasons: i) the simple form of the modes and the transmission and reflection coefficients allows an, albeit limited, analytical treatment, an attempt which is impossible for the higher modes, and ii) from this contribution 
we can draw some conclusions about the total probability. \par
If instead of a single localized detector we would use a detector somehow smeared on the surface of a spherical thin-shell, radially falling inwards towards the WH, then the s-wave contributions would represent the complet probability.
More exactly, there should be a coherent ensemble of detectors distributed evenly across the surface of a sphere, such that the response of the system is obtained as the sum of the responses of each detectors. 
The corresponsing amplitude is written as:
\begin{equation}
A_0 (E,\tau) = \int\limits_{-\infty}^{\tau} d\tau\,e^{iE\tau}\ \varphi^{\sigma\, *}_{\omega 00} ({\bf x}(\tau), t(\tau)),
\end{equation}
where $\varphi^{\sigma}_{\omega 00}$ is the mode \eqref{eq:modes_explicit} with $l= m = 0$. Written out explicitly, the amplitude takes the following form:
\begin{eqnarray} \label{eq:amp1}
t<0 : \ \ A^+_0(E,t) & = & \int\limits_{-\infty}^{\tau} d\tau\, e^{iE\tau} \left(\,f^+_{\omega 0} (r)\, Y_{00}(\phi,\theta) \,e^{-i\omega t}\right)^* \\
&=& \sqrt{\frac{1}{4\pi}}\int\limits_{-\infty}^{\tau} d\tau \, N_{\omega0}\left( h_0^{(1)}(\omega r) - R^{\,*}_{\omega0}\, h_0^{(2)}(\omega r)\right)\,e^{i(E\,+\gamma\omega)\tau} \nn\\
&=& \sqrt{\frac{1}{8\pi^2\omega}} \int\limits_{-\infty}^{\tau} d\tau\,\left(-i\frac{e^{i\omega r}}{r} + \frac{e^{2ia\omega}}{1+ia\omega}\frac{ie^{-i\omega r}}{r}\right)\,e^{i(E\,+\gamma\omega)\tau} \nn\\
&=&\frac{ie^{ia\omega}}{\sqrt{8\pi^2\omega}}\int\limits_{-\infty}^{\tau} d\tau \left(\frac{e^{i(E\,+\gamma\omega(1\,+\,v))\tau}}{(1+ia\omega)(a-\gamma v\tau)} - \frac{e^{i(E\,+\gamma\omega(1\,-\,v))\tau}}{a-\gamma v\tau}\right) \nn
\end{eqnarray}
\begin{eqnarray} \label{eq:amp2}
t>0 : \ \ A^+_0(E,t) & = & A^+_0(E,0^-) + \int\limits_{0}^{\tau} d\tau\, e^{iE\tau} \left(\,f^+_{\omega 0} (r)\, Y_{00}(\phi,\theta) \,e^{-i\omega t}\right)^* \\
& = & A^+_0(E,0^-)+\sqrt{\frac{1}{4\pi}}\int\limits_{0}^{\tau} d\tau  N_{\omega0}\left( T^{\,*}_{\omega0}\, h_0^{(2)}(\omega r)\right)e^{i(E\,+\gamma\omega)\tau} \nn\\
&=& A^+_0(E,0^-)-\sqrt{\frac{1}{8\pi^2\omega}} \int\limits_{0}^{\tau} d\tau\left(\frac{ia\omega\, e^{2ia\omega}}{1+ia\omega}\,\frac{ie^{-i\omega r}}{r}\right)e^{i(E\,+\gamma\omega)\tau} \nn\\
&=&A^+_0(E,0^-) - \frac{ia\omega}{1+ia\omega}\frac{ie^{ia\omega}}{\sqrt{8\pi^2\omega}}\int\limits_{0}^{\tau} d\tau \left(\frac{e^{i(E\,+\,\omega(1\,+\,v))\tau}}{a+\gamma v\tau}\right), \nn
\end{eqnarray}
and similarly for the second set of modes with $\sigma = -$. \par
The partial probability is obtained by summing the contribution of the two sets of modes and integrating over all frequencies:
\begin{equation} \label{eq:prob_wh}
\mathcal{P}_0 (E,t) = \int\limits_{0}^{\infty} d\omega\,\left( \,\vert A^+_0\vert^{\,2} + \vert A^-_0 \vert ^{\,2}\right)
\end{equation}
\par
The integral over frequencies in \eqref{eq:amp1} and \eqref{eq:amp2} can not be evaluated analytically, because of the complicated frequency depence of the transmission and reflection coefficients. We continue with 
a numerical and graphical analysis of the probability \eqref{eq:prob_wh}. 
\begin{figure} 
\centering
\includegraphics[width=0.7\linewidth]{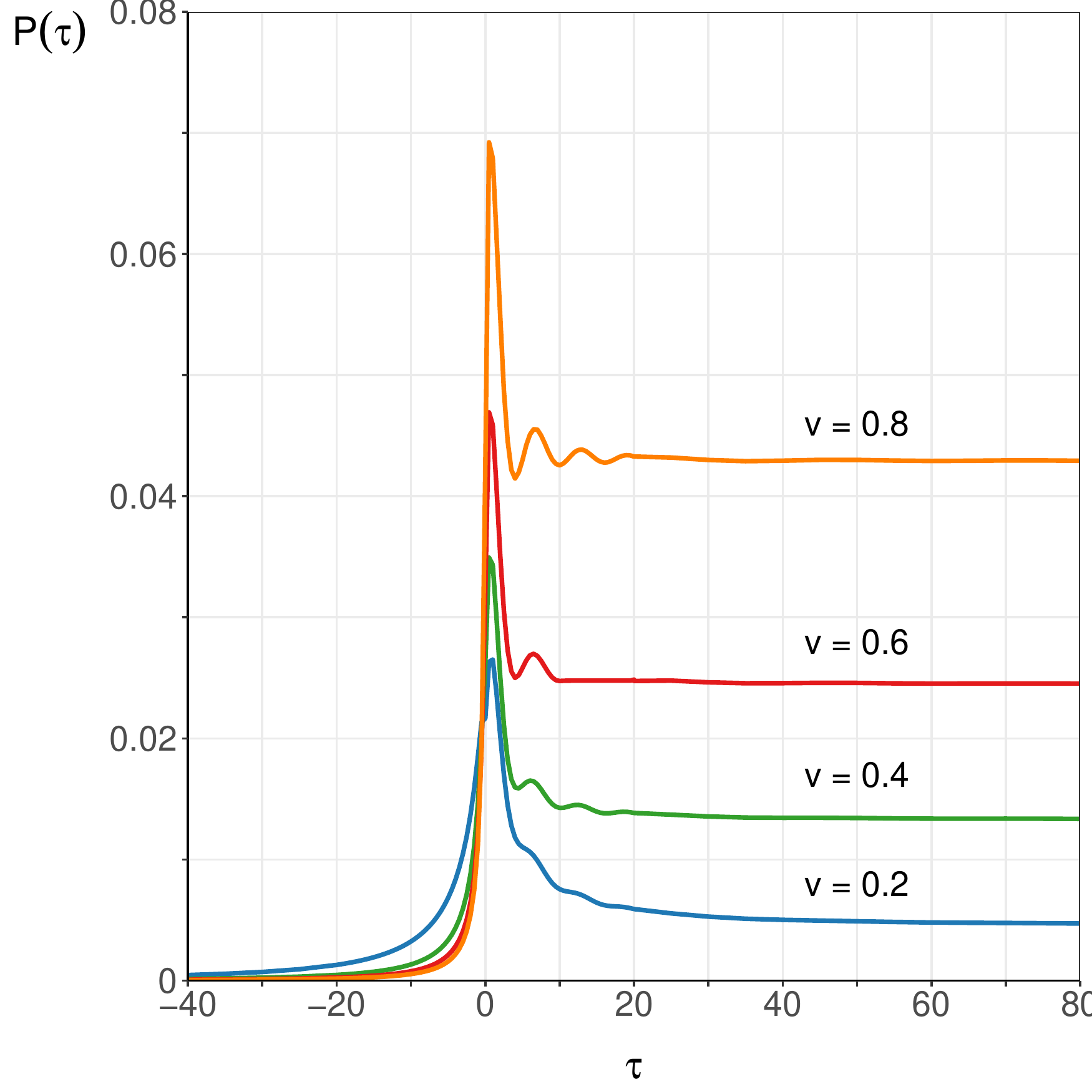}
\caption{\emph{ The contribution of the the $l=0$ (s-wave) mode to the total transition probability of the detector moving on the inertial trajectory \eqref{eq:trajectory_wh} through the wormhole spacetime described by the metric \eqref{eq:wh_metric}.}} 
\label{fig:4.5}
\end{figure}

The probability \eqref{eq:prob_wh} is represented in Fig.~\ref{fig:4.5}. As the detector zooms in from infinity, the probability starts to grow, peeking around the wormhole throat at $r = a$. After the detector passes through the wormhole, 
the probability starts decreasing, accompanied by spurious oscillations. As we progress to larger times/distances the oscillations get damped, and the probability stabilizes to a finite, non-zero value. \par

Compare this result with the response of the detector on an analogous trajectory in Minkowski space. There, the probability must vanish as we go towards $\tau \rightarrow \infty$. In the wormhole case the probability behaves 
similarly with the case when the inertial detector is perturbed by either switching on the coupling or having a discontinuity in the trajectory. 
As we have seen from Fig.\ref{fig:4.2}, after the detector leaves the wormhole the oscillations are quickly damped and the probability goes to a constant value. With this in mind,
we can approximate the probability at large enough times as equal to it's asymptotic value at $\tau \rightarrow \infty$. In this case the amplitude can be written with the help of the exponential integral function:
\begin{eqnarray} 
Ei(z) = \int\limits_{-\infty}^{z} \frac{e^t}{t} dt
\end{eqnarray}
Using the notation:
\begin{equation}
\alpha_{\pm} = (E + \gamma \omega (1\pm v) ) \frac{a}{\gamma v},
\end{equation}
the amplitude can be written as:
\begin{eqnarray}
T_1 &=& e^{i\frac{\alpha_\text{\small -}a}{v\gamma}}\left[\pi + iEi\left(-i\alpha_\text{\small -}a/v\gamma\right)\right] \\
T_2 &=& e^{i\frac{\alpha_\text{\tiny +}a}{v\gamma}}\left[\pi + iEi\left(-i\alpha_\text{\tiny +}a/v\gamma\right)\right] \\
T_3 &=& e^{-i\frac{\alpha_\text{\tiny +}a}{v\gamma}}\left[-\pi + iEi\left(i\alpha_\text{\tiny +}a/v\gamma\right)\right] \\
 A_0^+(E) &=& \frac{e^{i\omega a}}{\sqrt{8\pi^2\omega}}\left( T_1 - \frac{T_2 - i\omega a T_3}{1+i\omega a}\right),
\end{eqnarray}
and similarly the $\sigma = -$ modes.\par 

\subsection{Higher multipole contributions}
In what follows we look at how the probability changes as we sum over the contribution of the higher order modes ($l > 0$). The probability can be written as: 
\begin{equation}
\mathcal{P} (E,\tau)\ = \ \mathcal{P}_{l=0}(E,\tau)\ +\ \sum\limits_{l=1}^{\infty} \mathcal{P}_l(E,\tau)\ ,
\end{equation}
where the first terms is the s-wave contribution, analized in the preceding section, while the second term represents the effect of higher order modes. Due to the complicated form of the modes, and of the reflection and transmission
coefficients, we can not proceed through analytical methods. Thus, we must perform the sum numerically. The first step is to use complex-plane integration to make the integrals more convergent. The contour has to be closed in the
upper half of the imaginary plane. The amplitudes thus attain the following form:
\begin{eqnarray}
t<0&:& \     \ A^{(\pm)}_{\omega l}(E,\tau)=-ie^{iE\tau}\int\limits_{0}^{\infty} ds\ e^{-Es}(\varphi^{(\pm)}_{\omega l}(a-\gamma vt-i\gamma vs))^* \\
t>0&:& \     \ A^{(\pm)}_{\omega l}(E,\tau)= A^{(\pm)}_{\omega l}(E,0_-)+i\int\limits_{0}^{\infty}ds\ e^{-Es}(\varphi^{(\pm)}_{\omega l}(a+i\gamma vs))^*\\
&&\                        \ \ \ \ \ \ \ \ \ \ \ \ \ \ \      \ -ie^{iE\tau}\int\limits_{0}^{\infty}dt\ e^{-Es}(\varphi^{(\pm)}_{\omega l}(a+\gamma vt+i\gamma vs))^* \nn
\end{eqnarray}
\par
Next, we define a renormalized probability by subtracting the appropriate Minkowskian contribution. We have seen in the previous section that total and in most cases even the partial probabilities diverge for an inertial detector
in Minkowski space. Even though in the wormhole case, due to the non-vanishing radius of the wormhole, the arguments of the mode functions never reach the origin, and thus some of the divergences do not appear. Still, a part of the
resulting probability may just be a spurious contribution. Thus, in order to obtain the response of the detector generated purely by the presence of the wormhole, we subtract the appropriate Minkowskian contributions. The renormalization
is done at the level of the partial probabilities, for each mode in part:
\begin{equation}
 \mathcal{P}^\text{ren}_{\omega l} = \mathcal{P}_{\omega l} - \mathcal{P}^\text{M}_{\omega l},
\end{equation}
where $\mathcal{P}_{\omega l}$ represents the partial probability in the actualy WH spacetime, while $\mathcal{P}^\text{M}_{\omega l}$ is the analogous probability for a detector  moving with the same velocity on a linear radial
inertial trajectory, evaluated at the same instance of time. Basically, by subtracting the flat space contribution, we are calculating the correction term $\Delta P_{\omega l}$ in \eqref{eq:prob_wh_Mink+WH}, with the temporal integral having the upper
limit $\tau$ instead of future infinity. \par
\begin{figure} 
\centering
\includegraphics[width=0.7\linewidth]{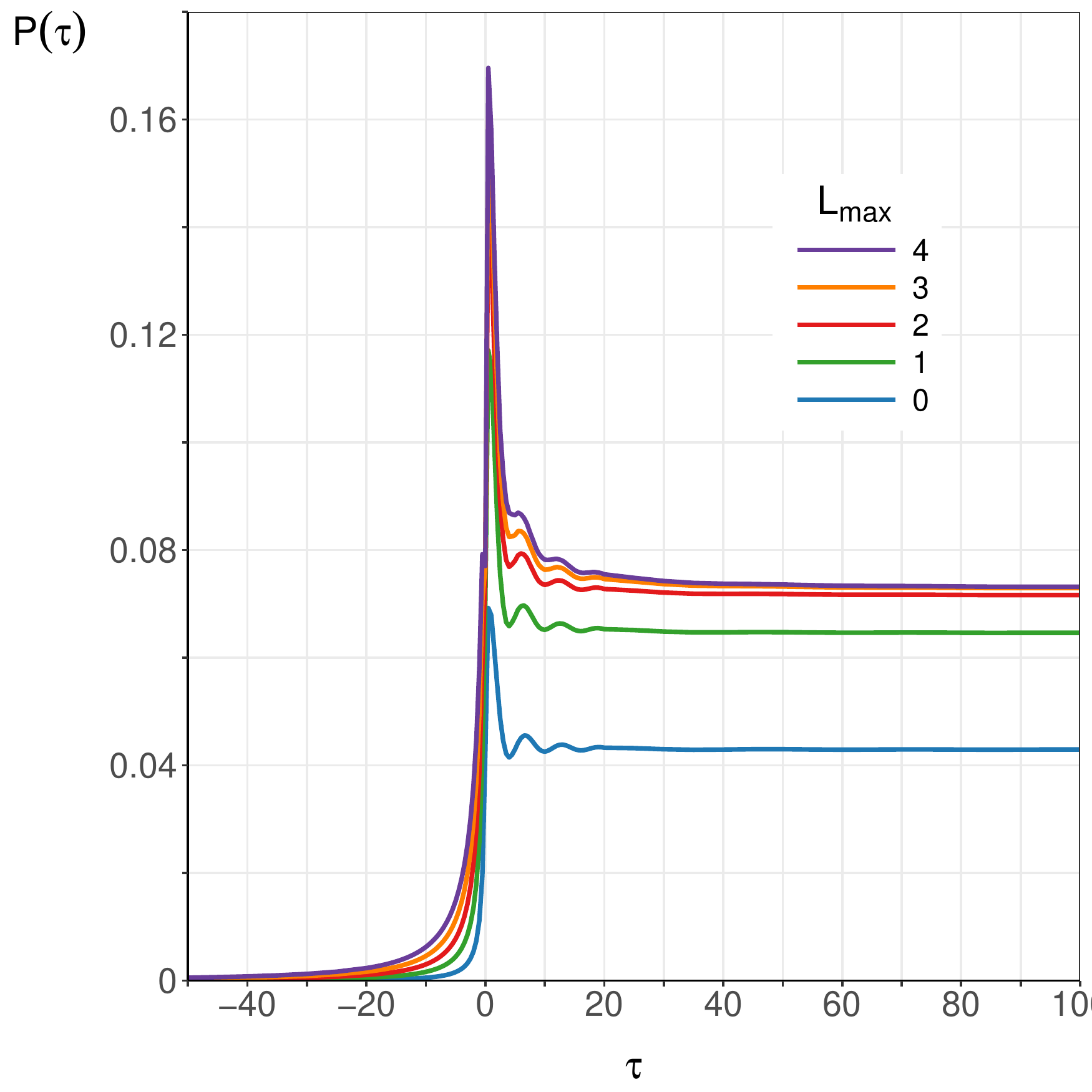}
\caption{\emph{ The effect of summing over higher order modes ($l > 0$) to the total transition probability. The different curves represent different values of $\text{L}_\text{max}$, up to which the summation is performed.
The fixed parameters are $E = 1$, $a = 1$, $v = 0.8$. }} 
\label{fig:5.5}
\end{figure}

The integration over frequencies is performed up to a value $\omega_\text{max}$. The required value is inversely proportional to the distance from the wormhole, and we optimize the numerical integration accordingly. Furthermore, for modes
with $\omega a \ll 1$ we have $T_{\omega l} \simeq 1$ and $R_{\omega l} \simeq 0$. Summing over both sets of modes (+ and -) we find that the contribution to the partial probabilities is equivalent to that in Minkowski space. Thus,
in this case we approximate the exact modes with the Minkowskian modes. This guarantees the elimination of possible large numerical errors in this domain, the renormalized partial amplitudes vanishing in this case.   
Finally, we must perform the summation over the quantum number l. In practice the sum has to be truncated at a finite value $L_\text{max}$. The convergence is very slow and thus many terms have to be taken into account. The best way 
to proceed is to fix an error threshold, for example $1\%$, and consider that convergence has been attained after the errors decrease below this value. In Fig.~\ref{fig:5.5} we have represented the cumulated contributions from the first 
5 modes. We see that at larger future times, after the tranzitory oscillations have become neglijible, the probability converges quite fast, a value of $L_\text{max} = 4$ being sufficient for convergence. Close to the throat of the wormhole, 
however the contributions of the higher multipoles are large, and the sum has to be extended to large values of $L_\text{max}$. 

 \chapter{Conclusion}
 In the present thesis we have investigated the Unruh effect in a spacetime containing a wormhole, described by the metric \eqref{eq:wh_metric}, by using the Unruh-deWitt monopole detector model. In the first chapter we have 
 reviewed the basic ingredients for quantifying fields on curved backgrounds along with the theory of particle detector. In the second chapter we have used the Unruh-deWitt detector to investigate the response of the detector on 
 various inertial and accelerated trajectories, recovering the well known results from the literature. We have also analysed trajectories with sudden variations in the velocity or the acceleration and scenarios where 
 the detector is coupled on at a finite time. In both situations the field becomes perturbed and the transition probability of the detector has strong variations, leading to departures from the expected results. In the forth 
 chapter we looked at detectors moving 
 on trajectories in a thin-shell wormhole spacetime. We have analysed in detail the reponse of the inertial detector moving radially through the wormhole. We found that the transition probability has the following structure:
 \begin{itemize}
  \item in the distant past, the probabality is very small, vanishing in the infinite past limit 
  \item as the detector nears the wormhole the probability starts rising, peaking around the throat of the WH
  \item after it crosses onto the other side, the probability starts decreasing, followed by a period of tranzitory oscillations
  \item as the detector departs, the oscillations are damped and the probability stabilizes to its final, nonzero value
 \end{itemize}
 Our results show that the Unruh-deWitt-type particle detector models are valuable tools for investigating the state of quantum fields in backgrounds with curvature and topological features. In the absence of a complete theory of
 quantum gravity, any new insights or contributions, no matter how small,  to our understanding of phenomena linking quantum physics and gravity, represent important stepping stones in our road towards understanding nature at the most
 fundamental level. 

\ \\ \ \\ \ 
\\ \ 
{\bf Acknowledgements}.\\ We thank Nistor Nicolaevici for guidance during the elaboration of this thesis.
\appendix
\chapter{Reflection and Transmission coefficients} \label{AppendixA}
For completeness, we list here the first few coefficients.
\begin{eqnarray}
R_{\omega\, 0}&=& \frac{e^{-2i\omega a}}{1-i\omega a},
\nn
\\
T_{\omega\, 0}&=&i\omega a \frac{e^{-2i\omega a}}{1-i\omega a}.
\nn
\end{eqnarray}
\begin{eqnarray}
R_{\omega 1}& =&
\frac{2+(a\omega)^2}{2-2ia\omega-a^2\omega^2} \,
\frac{e^{-2ia\omega}}{1-ia\omega},
\nn
\\
T_{\omega\, 1}&=&
\frac{i(a\omega)^3}{2-2ia\omega-(a\omega)^2} \,
\frac{e^{-2ia\omega}}{1-ia\omega}.
\nn
\end{eqnarray}
\begin{eqnarray}
R_{\omega 2} &=& \frac{27+6(a\omega)^2+(a\omega)^4}{9-9ia-4(a\omega)^2+i(a\omega)^3}
\,
\frac{e^{-2ia\omega}}{3-3ia\omega-(a\omega)^2},
\nn
\\
T_{\omega 2}& =&\frac{i(a\omega)^5}{9-9ia\omega-4(a\omega)^2+i(a\omega)^3}\,
\frac{e^{-2ia\omega}}{3-3ia\omega-(a\omega)^2}\,.
\nn
\end{eqnarray}
It is easy to see that these obey the relation:
\begin{equation}
 \vert R_{\omega\, l} \vert^2 + \vert T_{\omega\, l}\vert^2 = 1
\end{equation}

\bibliographystyle{plain}
\bibliography{byblos}
\end{document}